\documentclass[reprint, amsmath, amssymb, aps, pra]{revtex4-1}

\usepackage{subfigure}

\usepackage{amsmath}

\usepackage{graphicx}% Include figure files

\graphicspath{{images/}}

\usepackage{dcolumn}% Align table columns on decimal point

\usepackage{bm}% bold math

\usepackage[colorlinks=true,allcolors=blue]{hyperref}% add hypertext capabilities

\begin{document}

\title{Hidden symmetry and tunnelling dynamics in asymmetric quantum Rabi models}% Force line breaks with \\

\author{Zi-Min Li}
\affiliation{Department of Theoretical Physics, Research School of Physics, Australian National University, Canberra ACT, 2601, Australia}

\author{Murray T. Batchelor}
\email{batchelor@cqu.edu.cn}
\affiliation{Centre for Modern Physics, Chongqing University, Chongqing 400044, The People's Republic of China}
\affiliation{Department of Theoretical Physics, Research School of Physics, Australian National University, Canberra ACT, 2601, Australia}
\affiliation{Mathematical Sciences Institute, Australian National University, Canberra ACT, 2601, Australia}

\date{\today}% It is always \today, today,


\begin{abstract}
The asymmetric quantum Rabi model (AQRM) has a broken $\mathbb{Z}_2$ symmetry, with generally a non-degenerate eigenvalue spectrum. In some special cases where the asymmetric parameter is a multiple of the cavity frequency, stable level crossings typical of the $\mathbb{Z}_2$-symmetric quantum Rabi model are recovered, however, without any obvious parity-like symmetry. This unknown ``symmetry" has thus been referred to as hidden symmetry in the literature. Here we show that this hidden symmetry is not limited to the AQRM, but exists in various related light-matter interaction models with an asymmetric qubit bias term. Conditions under which the hidden symmetry exists in these models are determined and discussed. By investigating tunnelling dynamics in the displaced oscillator basis, a strong connection is found between the hidden symmetry and selective tunnelling.

\end{abstract}


\maketitle

\section{Introduction}
The quantum Rabi model (QRM) describes the simplest light-matter interaction, between a two-level atom and a single-mode bosonic field \cite{Rabi_1936,*Rabi_1937}. 
Despite its simplicity, the QRM is of fundamental importance in different areas of physics, particularly so in quantum optics \cite{Fox2006}. 
The past decade has seen the remarkable development of quantum technologies and experiments that give rise to realizations of the QRM in several advanced platforms. 
The most prominent example is circuit quantum electrodynamics (QED) \cite{Forn_D_az_2010,Yoshihara_2018,Blais2020}, where superconducting qubits play the role of artificial atoms and strongly interact with on-chip resonant circuits.

The Hamiltonian of the QRM reads ($\hbar=1$)
\begin{equation}\label{HQRM}
H_\text{R} = \dfrac{\Delta}{2}\sigma_z + \omega a^\dagger a + g \sigma_x \left( a + a^\dagger \right),
\end{equation}
where $\Delta$ is the level splitting of the qubit, $\omega$ is the field frequency, $a^\dagger$ and $a$ are the creation and annihilation operators, and $\sigma_{x,y,z}$ are the standard Pauli matrices. The QRM Hamiltonian commutes with the combined parity operator \cite{Xie_2017,Braak_2019}, i.e., $[H_R,P]=0$ with $P=\sigma_z\exp(\mathrm{i}\pi a^\dagger a)$. In other words, Eq.~(\ref{HQRM}) is invariant under the parity transformation $P^\dagger H_R P = H_R$. Thus the QRM possesses a $\mathbb{Z}_2$ symmetry \cite{Braak_2019}, which then permits spectral crossings of levels from different symmetry sectors. 

A well-known generalization of the QRM is obtained by adding an asymmetric qubit bias term in Eq.~(\ref{HQRM}), namely
\begin{equation}\label{haqrm}
H_\text{A} = \dfrac{\Delta}{2}\sigma_z + \dfrac{\epsilon}{2}\sigma_x + \omega a^\dagger a +  g\sigma_x ({a^\dagger} + a) . 
\end{equation}
This model is known as the asymmetric quantum Rabi model (AQRM) or biased QRM \cite{Braak_2011,Zhong_2014,Li_2015,Li_2016,Batchelor2015,Liu2017a,Mao_2018,Guan_2018,Xie2020}. 
The AQRM naturally appears in circuit QED systems, where the asymmetric term $\frac{\epsilon}{2}\sigma_x$ is the bias of the superconducting flux qubit, which can be tuned externally \cite{Forn_D_az_2010,Yoshihara_2018}. 

Obviously, the AQRM with nonzero $\epsilon$ does not commute with the parity operator. 
Indeed, Eq.~(\ref{haqrm}) is no longer invariant under the parity transformation, with
\begin{equation}\label{ParityTransformation}
P^\dagger H_A P = \dfrac{\Delta}{2}\sigma_z - \dfrac{\epsilon}{2}\sigma_x + \omega a^\dagger a +  g\sigma_x ({a^\dagger} + a). 
\end{equation}
As a result, the $\mathbb{Z}_2$ symmetry is broken. 
In fact, there is no explicit known symmetry in the AQRM. 

In general cases, the spectral levels of the AQRM do not cross, since they all belong to the same symmetry sector. 
Interestingly, however, when $\epsilon/\omega$ takes integer values, level crossings emerge in the AQRM, as if the $\mathbb{Z}_2$ symmetry is recovered \cite{Braak_2011}. 
Inspecting Eq.~(\ref{ParityTransformation}) shows that this is not the case. 
There must thus be some other type of parity-like symmetry in the AQRM, which is referred to as hidden symmetry \cite{Wakayama_2017,Kimoto_2020, Ashhab_2020}. 
Our aim here is to understand the hidden symmetry in a wider context. 

This paper is set out as follows.
In section \ref{HidenSymmetrySection}, we show the evidence for this hidden symmetry in the AQRM and 
then explore the same symmetry in other related models with a broken parity.  
We expect to get different conditions for $\epsilon$, which we call the $\epsilon$-condition. 
In particular, we explore how each term in the light-matter interaction Hamiltonians affects the hidden symmetry. 
Section \ref{SectiontunnellingDynamics} is devoted to a detailed discussion of the physical properties of the systems exhibiting hidden symmetry. 
Particular emphasis is given to the investigation of tunnelling dynamics in the relevant models within the displaced oscillator basis. 
We find a strong connection between the hidden symmetry and selective tunnelling. 
Concluding remarks, with an outlook on future research, are given in section \ref{SectionConclusion}.

\section{Hidden symmetry in related models}\label{HidenSymmetrySection}
\subsection{Asymmetric QRM}
The hidden symmetry of interest was first observed in the AQRM, and has been discussed by several authors \cite{Braak_2011, Gardas_2013, Zhong_2014, Li_2015, Wakayama_2017, Kimoto_2020, Ashhab_2020}. 

\begin{figure}[htbp]\centering                                              
	\subfigure[]{                
		\centering                                                 
		\includegraphics[width=0.465\linewidth]{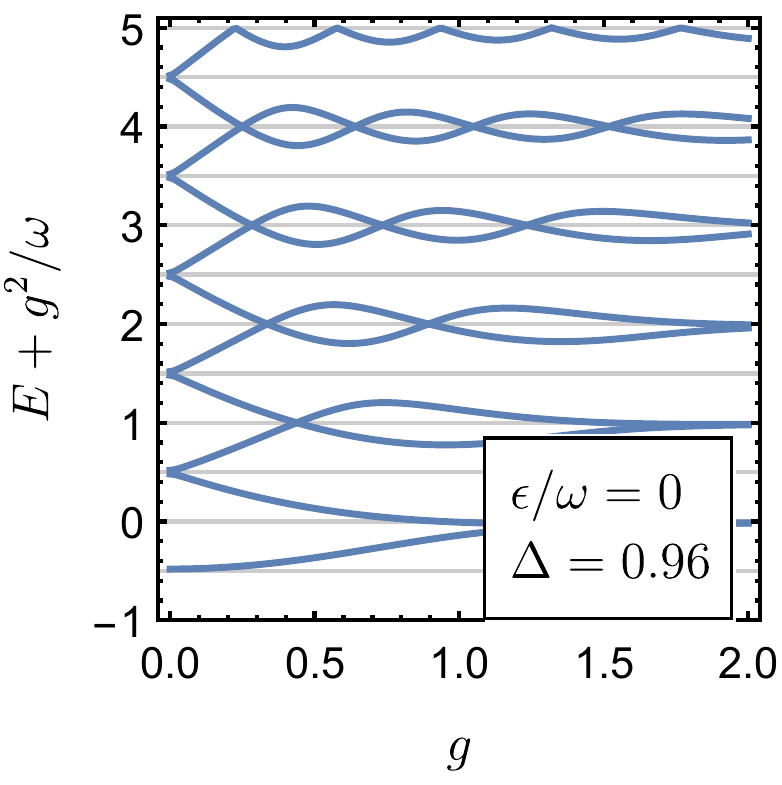}	\label{AQRMSpectrum1}  
	}   
	\subfigure[]{                
		\centering                                               
		\includegraphics[width=0.465\linewidth]{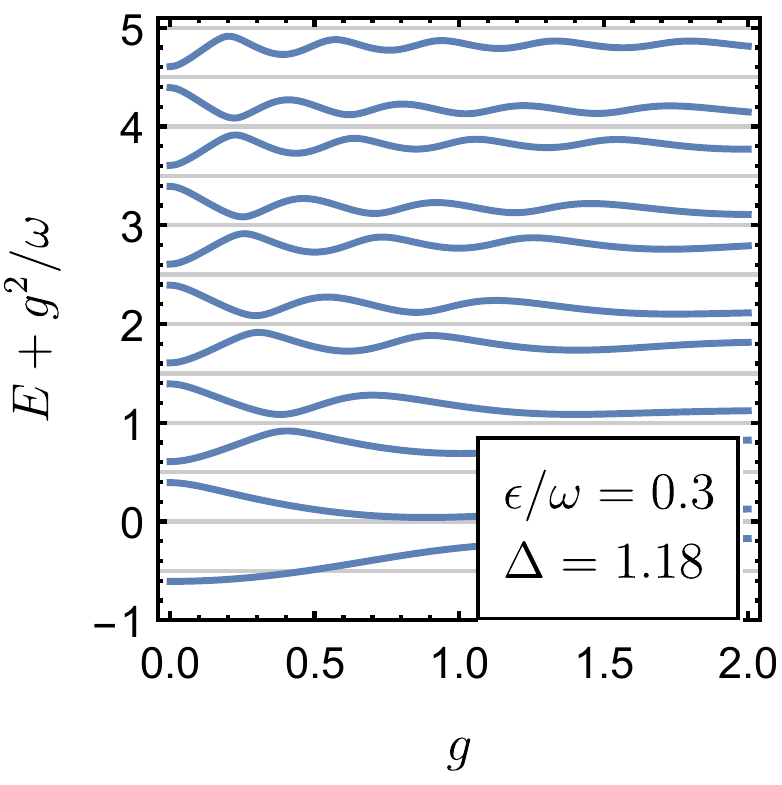}	\label{AQRMSpectrum2}          
	}
	\subfigure[]{                
		\centering                                                 
		\includegraphics[width=0.465\linewidth]{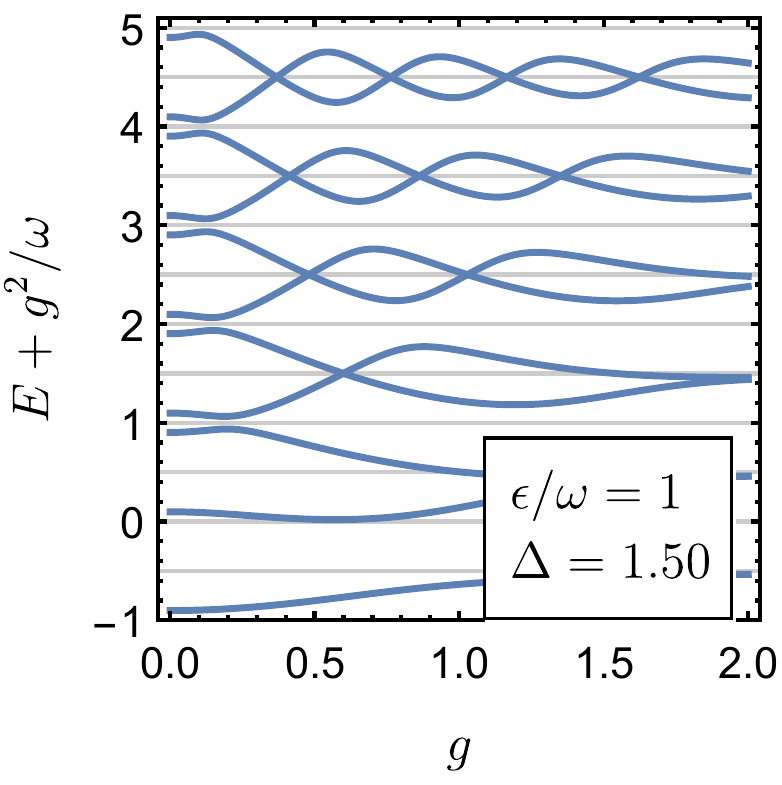}	\label{AQRMSpectrum3}        
	}   
	\subfigure[]{                
		\centering                                                 
		\includegraphics[width=0.465\linewidth]{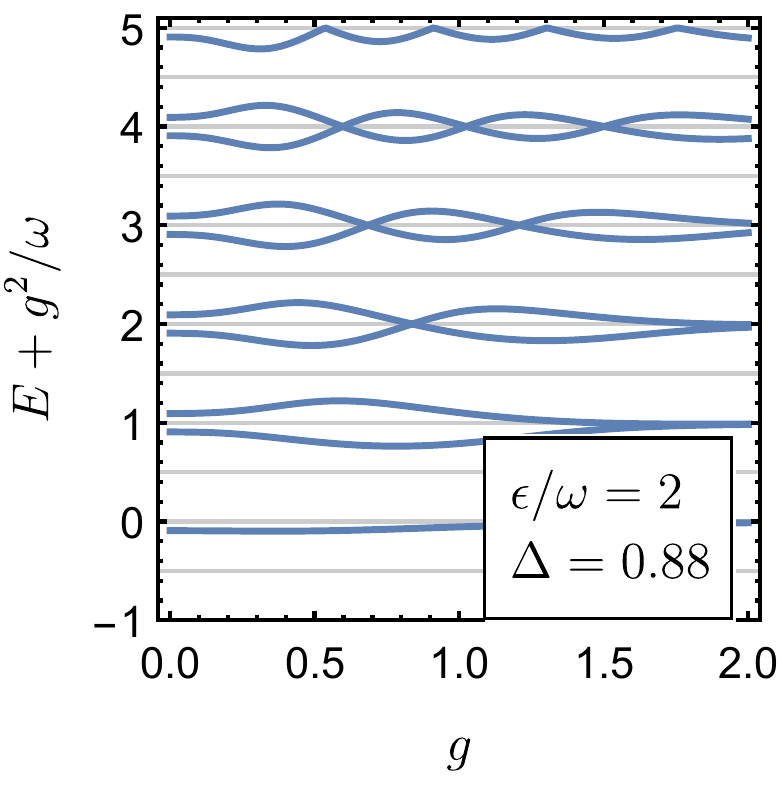}	\label{AQRMSpectrum4}        
	} 
	
	\caption{Spectrum of the asymmetric quantum Rabi model with respect to the coupling strength $g$ with  (a) $\epsilon/\omega=0$, (b) $\epsilon/\omega=0.3$, (c) $\epsilon/\omega=1$ and (d) $\epsilon/\omega=2$. The values of $\Delta$ are generated randomly in the range of $(0.5,2.5)$, to explicitly demonstrate that $\Delta$ is irrelevant to the existence of degeneracy. For clarity, the energies are rescaled with $E+g^2/\omega$.  } 
	\label{AQRMSpectrum}
\end{figure}

As pointed out in the Introduction, the $\mathbb{Z}_2$ symmetry in the QRM is broken by the $\epsilon$ term. 
Level crossings appear in the spectrum of the AQRM only if 
\begin{equation}\label{aqrmcondition}
\epsilon = n \omega, \quad n=1,2,3,\dots
\end{equation}
The spectrum exhibiting level crossings in Fig.~\ref{AQRMSpectrum1} belongs to the original QRM.
These crossing points are usually referred to as Juddian exceptional points \cite{Judd_1979}, which can be determined through polynomials in the system parameters \cite{Kus1985}. 
Fig.~\ref{AQRMSpectrum2} displays the most general case of the AQRM, where nonzero $\epsilon$ lifts the degeneracies and thus with no level crossings in the spectrum. 
In Fig.~\ref{AQRMSpectrum3} and \ref{AQRMSpectrum4}, where $\epsilon$ satisfies Eq.~(\ref{aqrmcondition}), level crossings appear again. 
These points have been proven to be real crossings \cite{Li_2015,Wakayama_2017}, and can be readily checked numerically \cite{Ashhab_2020}. 
Note that the value of $\Delta$ does not affect the existence of level crossings in the AQRM, as long as it is nonzero. 
To demonstrate this, we have chosen random values of $\Delta$ in the range of $(0.5,2.5)$ when generating the spectra in Fig.~\ref{AQRMSpectrum}. 

The simple $\epsilon$-condition (\ref{aqrmcondition}) coincides with the pole structure of the analytic solutions to the QRM, 
see Refs.~\cite{Braak_2011, Chen2012, Zhong_2013,Maciejewski_2014}. 
This observation can also be made in the spectrum, where, for example, exceptional energy baselines $E_n = n \omega +g^2/\omega + \epsilon/2$ and $E_{n+1} = (n+1) \omega +g^2/\omega - \epsilon/2$ coincide when $\epsilon = \omega$ \cite{Zhong_2014,Li_2015,Wakayama_2017}. 

In the following, we investigate level crossings in other related models, and we will see that the $\epsilon$-conditions therein are not as simple as 
Eq.~(\ref{aqrmcondition}). 
However, they may also be found through the pole structure of the analytic solutions.

\subsection{Asymmetric Rabi-Stark model}
\begin{figure}[t]\centering                                              
	\subfigure[]{                
		\centering                                                 
		\includegraphics[width=0.448\linewidth]{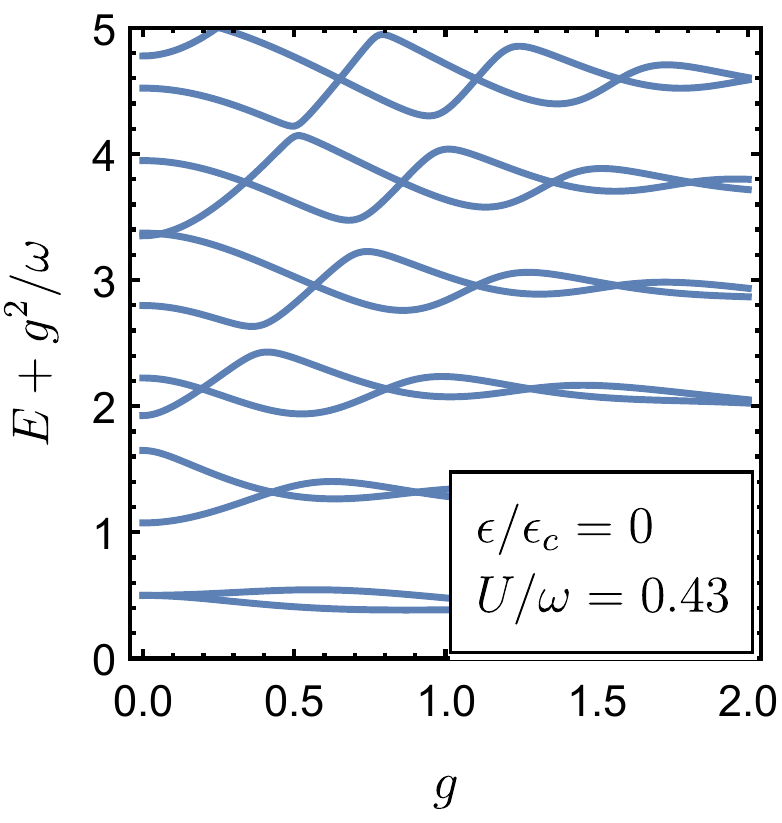}	\label{ARSMSpectrum1}        
	}   
	\subfigure[]{                
		\centering                                               
		\includegraphics[width=0.448\linewidth]{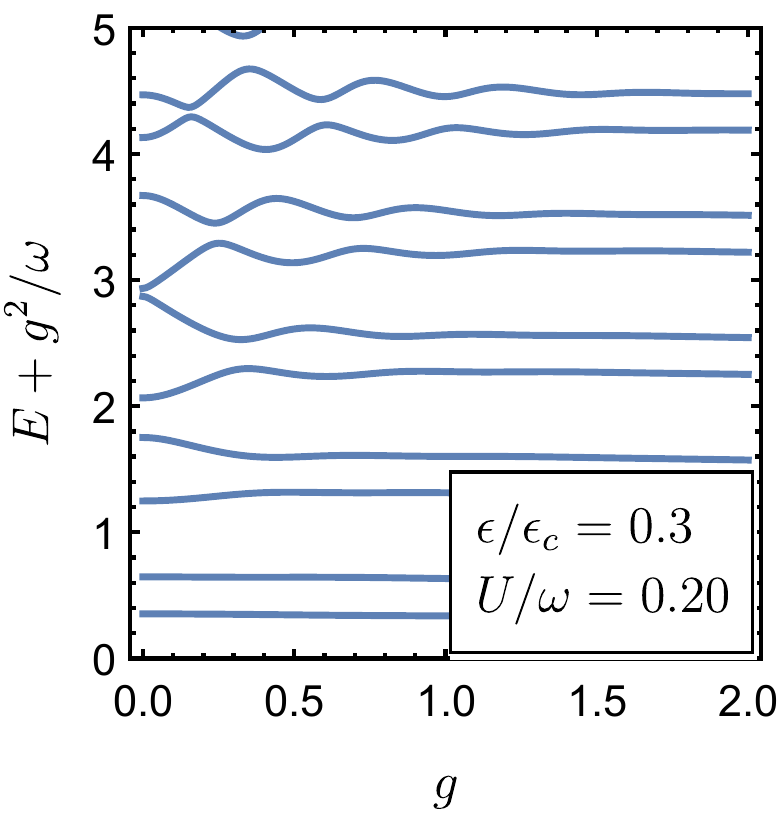}	\label{ARSMSpectrum2}          
	}
	\subfigure[]{                
		\centering                                                 
		\includegraphics[width=0.448\linewidth]{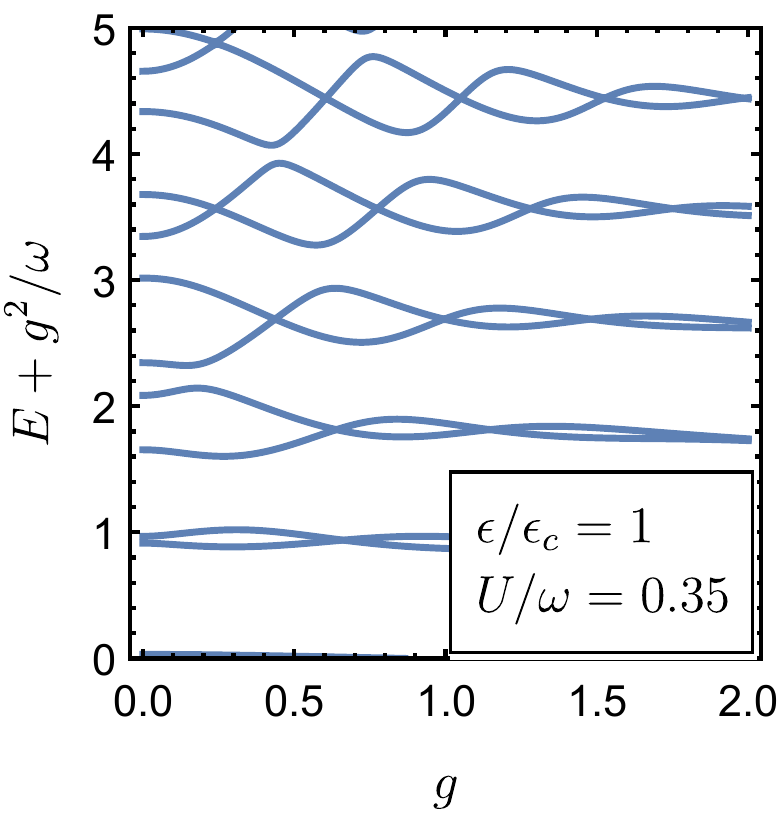}	\label{ARSMSpectrum3}        
	}   
	\subfigure[]{                
		\centering                                                 
		\includegraphics[width=0.448\linewidth]{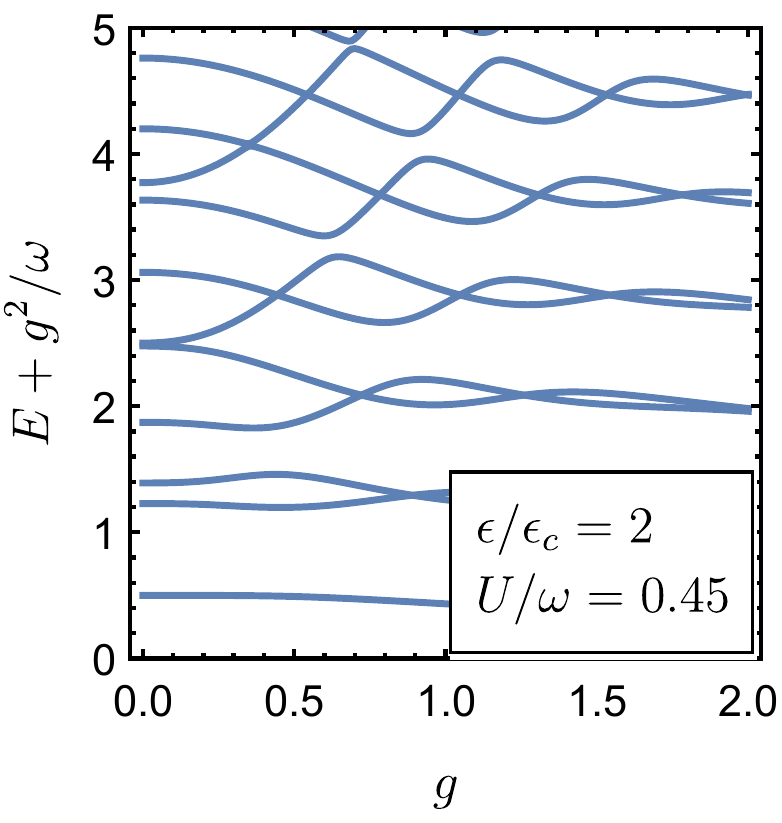}	\label{ARSMSpectrum4}        
	} 
	
	\caption{Spectrum of the asymmetric Rabi-Stark model with respect to the coupling strength $g$, with $\Delta=1$, $\omega=1$, and random values of $U/\omega$ in the range of $(0,0.75)$. The bias parameter has values (a)  $\epsilon/\epsilon_c=0$, (b) $\epsilon/\epsilon_c=0.3$, (c) $\epsilon/\epsilon_c=1$ and (d) $\epsilon/\epsilon_c=2$, where $\epsilon_c$ is determined through Eq.~(\ref{arsmcondition}). For clarity, the energies are rescaled with $E+g^2/\omega$. } 
	\label{ARSMSpectrum}
\end{figure}

The QRM with a Stark coupling term, usually called the Rabi-Stark model, has attracted recent interest \cite{Grimsmo_2013,*Grimsmo_2014,Eckle_2017,Xie_2019, Chen2020, Xie_2020, Cong_2020}. 
In the same manner as the AQRM, the asymmetric Rabi-Stark model (ARSM) is obtained by adding the asymmetric term,
\begin{equation}\label{harsm}
H_\text{RS} = \left(\dfrac{\Delta}{2}+ U a^\dagger a\right)\sigma_z  + \omega a^\dagger a + g \sigma_x ( a^\dagger + a ) + \dfrac{\epsilon}{2}\sigma_x .   
\end{equation}
Here $|U/\omega|<1$ is required to avoid unphysical results \cite{Grimsmo_2013}.
As in the AQRM, Eq.~(\ref{harsm}) also has $\mathbb{Z}_2$ symmetry when $\epsilon=0$ and there are stable level crossings in the spectrum, as shown in Fig.~\ref{ARSMSpectrum1}. 
The general non-crossing case of the ARSM spectrum is shown in Fig.~\ref{ARSMSpectrum2} with $\epsilon/\epsilon_c=0.3$, where degeneracies are lifted and no level crossings exist. 

The hidden symmetry is observed when
\begin{equation}\label{arsmcondition}
\epsilon = n \epsilon_c,\quad \epsilon_c = \sqrt{\left(\omega - U\right)\left(\omega + U\right)}. 
\end{equation}
In Fig.~\ref{ARSMSpectrum3} and \ref{ARSMSpectrum4}, where the $\epsilon$-condition Eq. (\ref{arsmcondition}) is satisfied, 
the level crossings again appear. We have numerically verified that these are real crossings. 
Here again, to show that the existence of hidden symmetry does not rely on other parameters, we have chosen to generate $U$ randomly in the range $(0,0.75)$.

In fact the ARSM represents a family of models in which the field frequency is modified.  
We now consider a slightly different model, namely
\begin{equation}\label{hwrong}
H_\text{RS}' = H_A \pm U a^\dagger a\sigma_\pm\sigma_\mp , 
\end{equation}
in which the photon frequency is altered either in the spin-up or spin-down subspace. 
The hidden symmetry also exists in this model, with the corresponding $\epsilon$-condition one would expect, i.e.,
\begin{equation}\label{wrongcondition}
\epsilon = n \epsilon_c',\quad \epsilon_c' = \sqrt{\omega\left(\omega \pm U\right)}. 
\end{equation}
There is a high similarity between the Hamiltonians in Eq.~(\ref{haqrm}) and Eq.~(\ref{hwrong}) and their spectra, 
so we do not include these figures here. 

Up to this point, we have seen that the modification in frequency affects the hidden symmetry and the $\epsilon$-condition.

\subsection{Anisotropic AQRM}
The anisotropic AQRM \cite{Tomka_2014, Xie_2014} consists of different coupling strengths for two kinds of light-matter interaction terms, i.e., the energy conserving co-rotating terms and the energy non-conserving counter-rotating terms. 
Exploring the anisotropic AQRM provides some intuition about how the interaction terms affect the hidden symmetry. 
The Hamiltonian of the anisotropic AQRM is
\begin{equation}\label{anisotropicQRM}
\begin{aligned}
H_\text{an} =& \omega a^\dagger a + \dfrac{\Delta}{2}\sigma_z + \dfrac{\epsilon}{2}\sigma_x \\
&+ g_1(\sigma_-{a^\dagger} + \sigma_+a) + g_2(\sigma_+{a^\dagger} + \sigma_- a). 
\end{aligned}
\end{equation}
When $\epsilon=0$, the anisotropic QRM interpolates between the Jaynes-Cummings model ($g_2=0$) \cite{Jaynes_1963} with $U(1)$ symmetry and the QRM ($g_1=g_2$) with $\mathbb{Z}_2$ symmetry.
For $g_1=0$, we obtain the model we shall call the anti-Jaynes-Cummings model. 
Here we restrict ourselves to the case where both $g_1$ and $g_2$ are nonzero, and $g_2 = \lambda g_1, \lambda \ne 0$. 
The $\mathbb{Z}_2$ symmetry is broken if $\epsilon \ne 0$.

\begin{figure}[t]\centering                                              
	\subfigure[]{                
		\centering                                                 
		\includegraphics[width=0.469\linewidth]{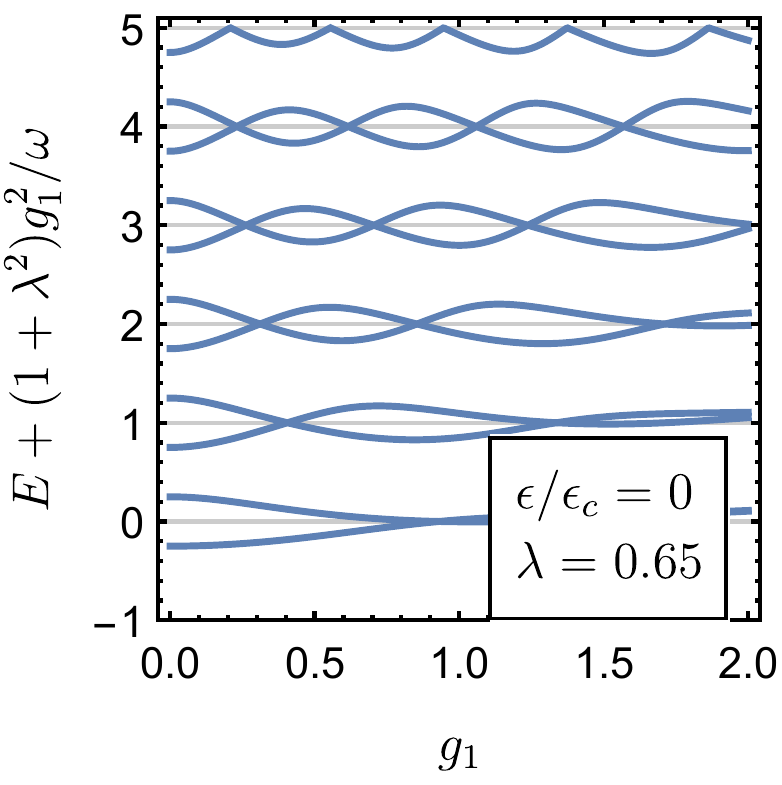}        
	}   
	\subfigure[]{                
		\centering                                               
		\includegraphics[width=0.469\linewidth]{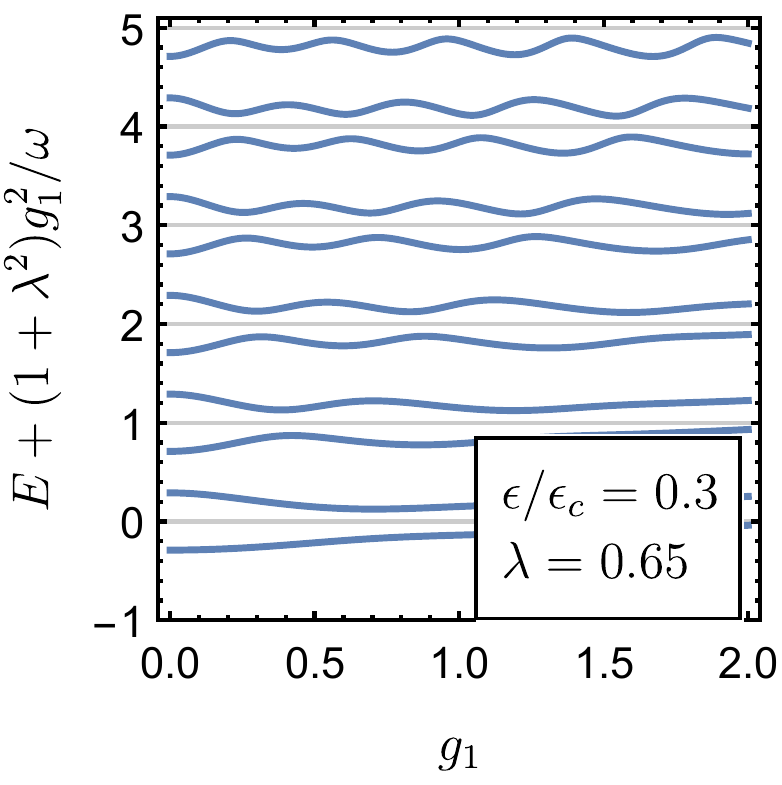}          
	}
	\subfigure[]{                
		\centering                                                 
		\includegraphics[width=0.469\linewidth]{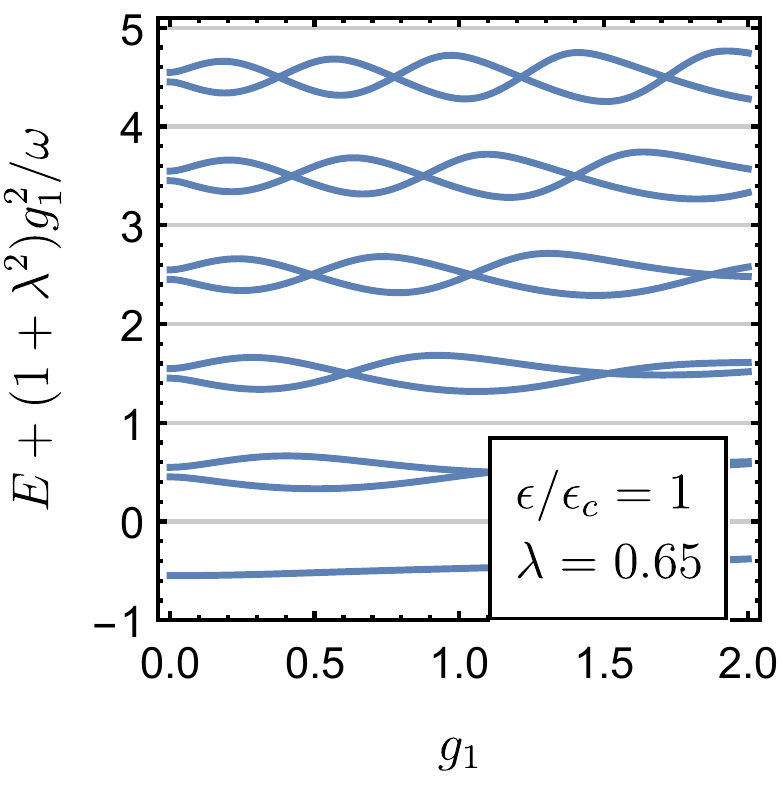}        
	}   
	\subfigure[]{                
		\centering                                                 
		\includegraphics[width=0.469\linewidth]{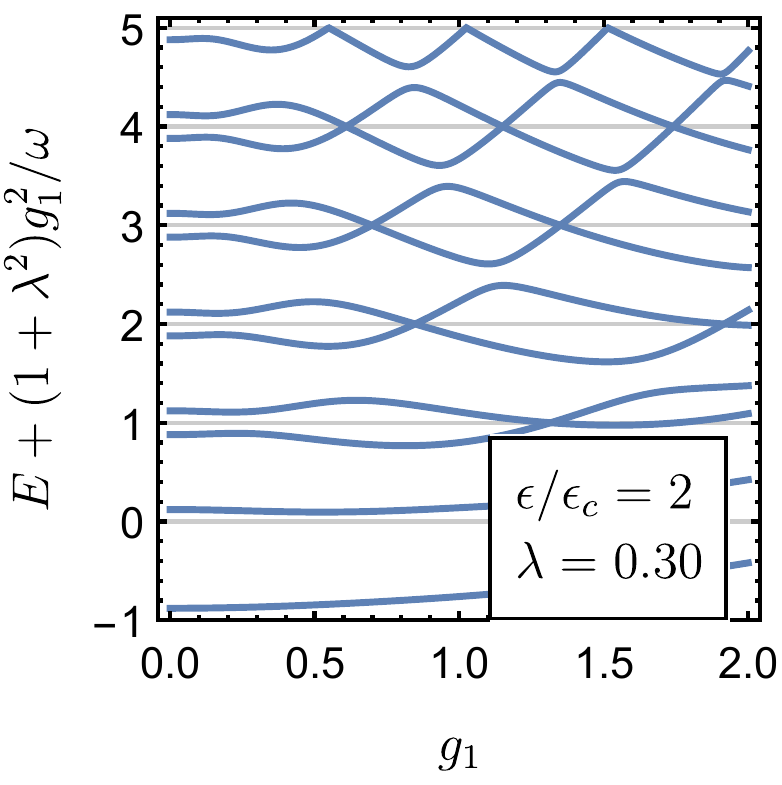}        
	} 
	
	\caption{Spectrum of the anisotropic asymmetric quantum Rabi model with respect to the coupling strength $g_1$, with $g_2 = \lambda g_1, \omega=1, \Delta=0.5$ and other parameter values as denoted in the figure. Here $\epsilon_c=\epsilon_c^{an}$ is determined through 
	Eq.~(\ref{anqrmcondition}). For clarity, the energies are rescaled with $E+(1+\lambda^2)g_1^2/\omega$. } 
	\label{AnisotropicQRMSpectrum}
\end{figure}

The hidden symmetry exists in the anisotropic AQRM defined in Eq.~(\ref{anisotropicQRM}), given that $g_2 = \lambda g_1$. 
By observing the pole structure in the analytic solutions \cite{Xie_2014}, we find the $\epsilon$-condition for this model to be
\begin{equation}\label{anqrmcondition}
\epsilon = n \epsilon_c^{an},\quad \epsilon_c^{an} =\dfrac{2\sqrt{\lambda}}{1+\lambda}\omega, 
\end{equation}
which reduces to Eq.~(\ref{aqrmcondition}) when $\lambda=1$. 

Fig.~\ref{AnisotropicQRMSpectrum} shows the spectrum for the anisotropic AQRM with various $\epsilon$.
The spectrum of the symmetric case with $\epsilon=0$ is similar to that of the original QRM. 
Level crossings are lifted with any  $\epsilon$ value that does not obey the $\epsilon$-condition (\ref{anqrmcondition}). 
When Eq.~(\ref{anqrmcondition}) is satisfied, on the other hand, the level crossings are present again. 

In the case where $g_1$ and $g_2$ are independent of each other, no hidden symmetry is found. 

\subsection{Other related models}
Until now we have seen that the hidden symmetry is affected by the modification of field frequency and anisotropy in the interaction terms. 
It is then natural to ask how other relevant parameters would affect the hidden symmetry. 
Here we discuss a few more related models. 
Since these models are much more complicated than the AQRM which simply considers a single mode and a single atom, we only aim to explore the existence of hidden symmetry in these models, without pursuing the corresponding $\epsilon$-conditions analytically. 

\subsubsection{Two-mode AQRM}
We now consider the two-mode AQRM \cite{Chilingaryan_2015}, whose Hamiltonian takes the form
\begin{equation}\label{twomodeqrm}
H_\text{tm} = \dfrac{\Delta}{2}\sigma_z +\dfrac{\epsilon}{2}\sigma_x + \sum_{i=1}^{2}\omega_i a_i^\dagger a_i + \sum_{i=1}^{2}g_i \sigma_x (a_i^\dagger + a_i) ,
\end{equation}
for which the $\mathbb{Z}_2$ symmetry exists when $\epsilon=0$. 

Observing that there are always crossings when $\omega_1 = \omega_2$, we conjecture there is an $|\omega_1-\omega_2|$ term in its $\epsilon$-condition. 
When $\omega_1 \ne \omega_2$, the existence of the hidden symmetry is unclear. 

We would like to point out that there is another form of the two-mode AQRM, in which the interaction term is written as $g\sigma_x (a_1^\dagger a_2^\dagger + a_1 a_2)$ \cite{Zhang2013}.
However, this alternative Hamiltonian has $U(1)$ symmetry \cite{Duan_2015}, rather than $\mathbb{Z}_2$. 

\subsubsection{Two-qubit AQRM}
Hidden symmetry in the two-qubit AQRM \cite{Zhang2013, Chilingaryan_2013, Peng_2014, Sun2019} is also worth considering.
Even though the atomic term $\Delta\sigma_z/2$ in the AQRM does not affect the existence of level crossings, we still seek to understand what role the qubit plays here. 
The Hamiltonian of interest reads
\begin{equation}\label{twoqubitQRM}
H_\text{tq} =   \sum_{i=1}^{2}\left(\dfrac{\Delta_i}{2}\sigma_{iz} + \dfrac{\epsilon_i}{2}\sigma_{ix}\right) 
+ \omega a^\dagger a + \sum_{i=1}^{2}g_i\sigma_{ix}(a^\dagger + a) ,
\end{equation}
which also possesses $\mathbb{Z}_2$ symmetry when $\epsilon_1=\epsilon_2=0$.

We find that only in the identical qubits case, i.e., when $\Delta_1 = \Delta_2, g_1 = g_2 $, does the level crossings appear with nonzero biases $\epsilon_1 = \epsilon_2$. 
This could be related to the permutation symmetry of identical qubits, as mentioned in \cite{Peng_2014}. 
However, we cannot rule out the possibility of the hidden parity-like symmetry that we emphasize in this article. 
Unlike for the previous models discussed, all the parameters in Eq.~(\ref{twoqubitQRM}) have effects on the existence of level crossings, which makes finding the $\epsilon$-condition through observation difficult. 

It is worth noting that the so-called ``dark states" \cite{Peng_2014} that exist in the original two-qubit QRM also appear in the asymmetric counterpart under the condition $\epsilon_1=\epsilon_2$, of course, also in the identical qubits case.

\section{Tunnelling dynamics}\label{SectiontunnellingDynamics}
Having demonstrated the existence of the hidden symmetry in various models, let us now concentrate on two cases in which the $\epsilon$-conditions are explicitly known: the AQRM (\ref{haqrm}) and the ARSM (\ref{harsm}). 
Although no corresponding constant of motion with physical interpretation has been found, it is still possible to understand the symmetry intuitively from some physical properties. 
Particularly, we consider the tunnelling dynamics in the displaced oscillator basis. 

\subsection{Displaced oscillator basis}
Following what has been adapted in the original QRM to derive the adiabatic approximation \cite{Irish_2005}, we can regard the part of the AQRM Hamiltonian 
\begin{equation}\label{hdo}
H_\text{do} =  \omega a^\dagger a +  g\sigma_x ({a^\dagger} + a) + \dfrac{\epsilon}{2}\sigma_x 
\end{equation}
as two displaced harmonic oscillators, with displacing directions determined by the two eigenvalues of $\sigma_x$. 
The $\frac{\Delta}{2}\sigma_z$ term in Eq.~(\ref{haqrm}) therefore describes the tunnelling between these two oscillators. 
The eigenstates of Eq.~(\ref{hdo}) are 
\begin{equation}\label{HdoEigenstates}
| n_\pm, \pm \rangle = |n_\pm\rangle \otimes |\pm\rangle, 
\end{equation}
with $|\pm\rangle$ satisfying  $\sigma_x|\pm\rangle = \pm| \pm \rangle$ and  $|n_\pm\rangle $ being the displaced Fock states, or the so-called generalized coherent states, namely
\begin{equation}\label{GeneralizedCoherentStates}
|n_\pm \rangle = \exp\left[\pm\frac{g}{\omega}(a-a^\dagger)\right]|n\rangle . 
\end{equation}
Here $|n\rangle$ with $n=0,1,2,\dots$ are the standard Fock states. 
The corresponding energy eigenvalues are given by 
\begin{equation}\label{HdoEigenvalues}
E_{n}^{\pm} = n \omega -\dfrac{g^2}{\omega} \pm \dfrac{\epsilon}{2}, 
\end{equation}
in which nonzero $\epsilon$ lifts the degeneracy and leads to asymmetry in the oscillator potentials, corresponding to the breaking of parity symmetry. 
While the eigenstates Eq.~(\ref{HdoEigenstates}) form a basis, the generalized coherent states themselves obey different orthogonal conditions,
\begin{equation}\label{GeneralizedCoherentStatesOrthogonality}
\langle m_\pm | n_\pm \rangle = \delta_{mn},\quad \langle m_\pm | n_\mp \rangle \ne 0. 
\end{equation}
In the following numerical computations, we will apply exact diagonalization to the Hamiltonian matrices expressed in the basis of 
Eq.~(\ref{HdoEigenstates}). 
Details can be found in Appendix. \ref{AppendixDisplacedOscillator}.

\begin{figure}[t]\centering      
	\subfigure[]{                
		\centering                                                 
		\includegraphics[width=0.45\linewidth]{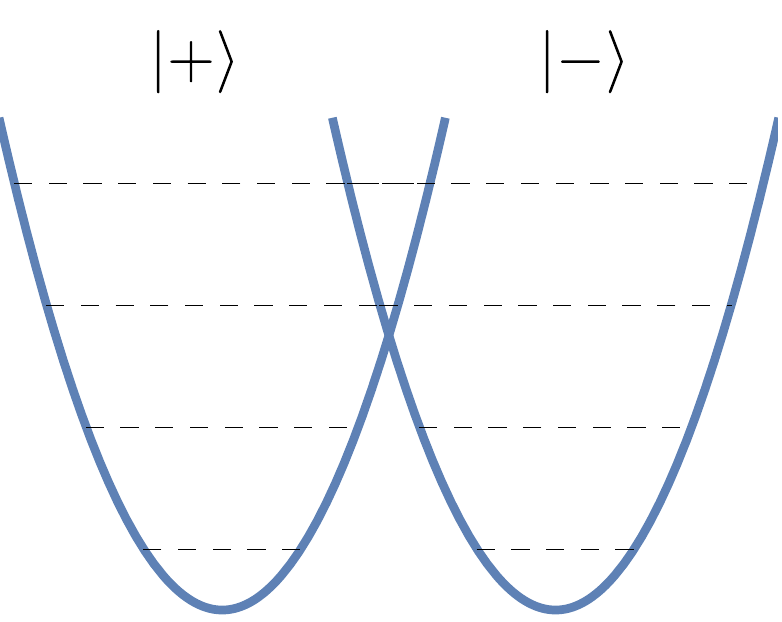}     
		\label{EffectivePotentialsQRM}   
	}    
	\subfigure[]{                
		\centering                                                 
		\includegraphics[width=0.45\linewidth]{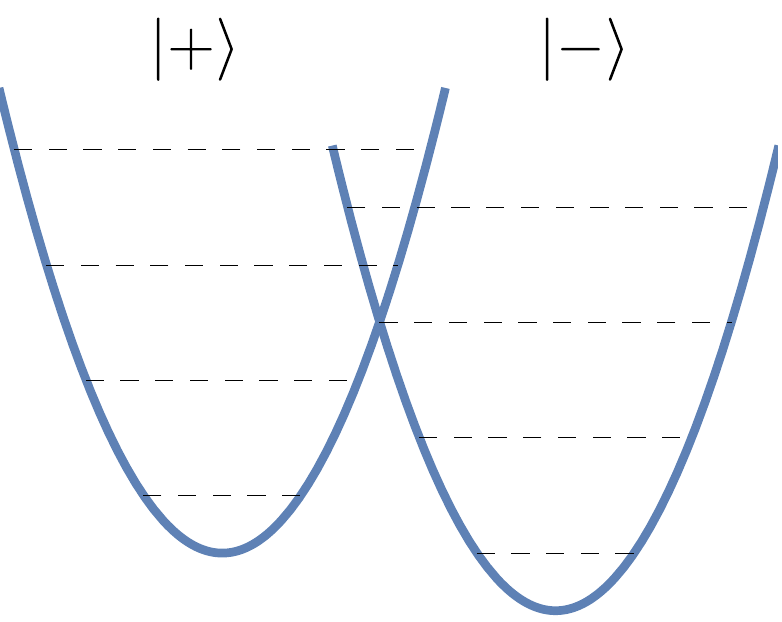}     
		\label{EffectivePotentials}   
	}   
	\subfigure[]{                
		\centering                                                 
		\includegraphics[width=0.9\linewidth]{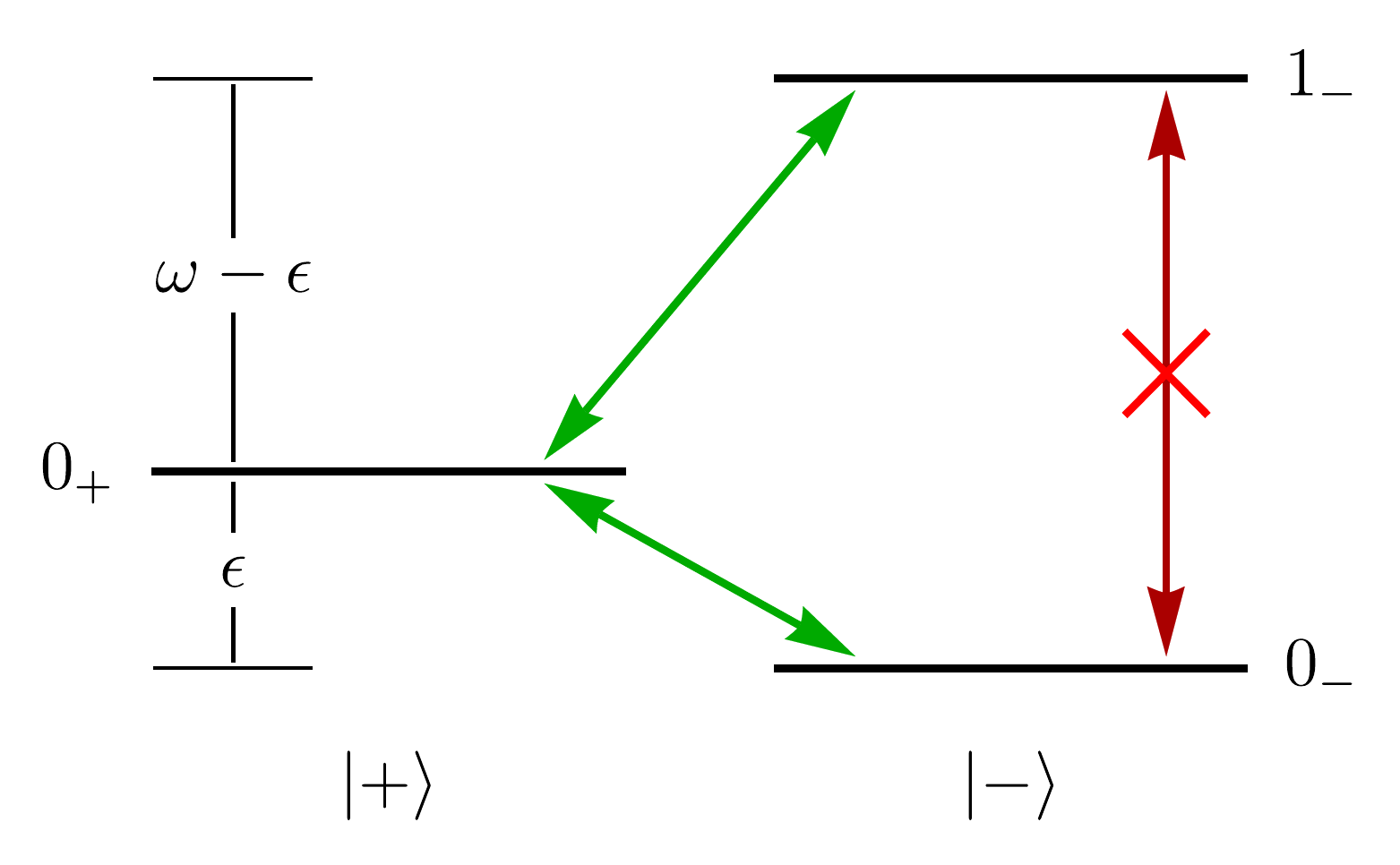}      
		\label{3Level}  
	}  
	\caption{(a) Schematic effective potentials of Eq.~(\ref{hdo}) for $\epsilon=0$. The potentials are symmetric, corresponding to the $\mathbb{Z}_2$ symmetry in the standard QRM. (b) Schematic effective potentials of Eq.~(\ref{hdo}) for nonzero $\epsilon$. The potentials are now asymmetric, corresponding to broken $\mathbb{Z}_2$ symmetry.  (c) Lowest three levels in (b). For $\Delta/\omega\ll 1$, the tunnelling process in the AQRM can be simplified into a three-level system. The AQRM Hamiltonian matrix elements admit tunnelling between different oscillators and forbid transitions within the same oscillator. } 
\end{figure}

The effective potentials and energy levels of Eq.~(\ref{hdo}) are  displayed schematically in Fig.~\ref{EffectivePotentialsQRM} and Fig.~\ref{EffectivePotentials}. If $\epsilon=0$, the energy levels of Eq.~(\ref{hdo}) are doubly degenerate, and the two oscillators in Fig.~\ref{EffectivePotentialsQRM} are symmetric with respect to the $E$-axis. 
Nonzero $\epsilon$ breaks this degeneracy and shifts the oscillators in opposite directions along the energy axis (up or down), as in 
Fig.~\ref{EffectivePotentials}.  
When $\epsilon = m \omega$, which coincides with the $\epsilon$-condition of the AQRM in Eq.~(\ref{aqrmcondition}), levels $E_{n}^+$ are degenerate with $E_{n+m}^-$, while lower levels of the $|-\rangle$ oscillator stay non-degenerate. 

\subsection{Tunnelling Dynamics in the AQRM}
Having established the displaced oscillator basis, we now consider the tunnelling dynamics \cite{Irish_2014} of the AQRM. 
The tunnelling term $H_t^a=\frac{\Delta}{2}\sigma_z$ in the AQRM Hamiltonian (\ref{haqrm}) ensures that transitions can only take place between different oscillators. 
Moreover, in the parameter regime $\Delta/\omega\ll 1$, numerical observation shows that tunnelling and reflections between levels with large energy gaps are negligible. 
The tunnelling process in this parameter regime can therefore be safely approximated as a three level transition problem, as depicted in Fig.~\ref{3Level}. 
Suppose the system is in the state $| 0_+, + \rangle$ at time $t=0$. 
The tunnelling matrix elements calculated in Appendix \ref{AppendixDisplacedOscillator} admit the evolution towards levels $|0_-,-\rangle$ and $| 1_-, - \rangle$, and forbid the transition within the same oscillator. 

%We assume the system is in the state $| 0_+, + \rangle$ at time $t=0$.
%Our target is to investigate how the system state evolves with time. 
%The levels 1,2,3 in Fig. \ref{3Level} then represent states $|0_-,-\rangle, | 0_+, + \rangle$ and $| 1_-, - \rangle$, respectively. 
According to standard time-dependent perturbation theory, the tunneling frequencies of the processes $|0_+,+\rangle \leftrightarrow | 0_-, - \rangle$ and $| 0_+, + \rangle \leftrightarrow | 1_-, - \rangle $ are calculated as
\begin{equation}\label{tunnellingFrequency}
\begin{aligned}
\omega_{00} = \dfrac{1}{2}\sqrt{\delta_{00}^2 + 4\Omega_{00}^2} ,\quad \omega_{01} = \dfrac{1}{2}\sqrt{\delta_{01}^2 + 4\Omega_{01}^2}. 
\end{aligned}
\end{equation}
Here $\delta_{00}=\epsilon$ and $\delta_{01}=\omega-\epsilon$ are the level gaps as displayed in Fig.~\ref{3Level}.
$\Omega_{00} = \frac{\Delta}{2}\langle 0_- | 0_+ \rangle$ and $\Omega_{01} = \frac{\Delta}{2}\langle 1_- | 0_+ \rangle$ are the corresponding tunnelling matrix elements (see detailed calculations in Appendix \ref{AppendixtunnellingFrequency} ). 

When $\epsilon=0$, level $| 0_+, + \rangle$ and $|0_-,-\rangle$ are degenerate, and the tunnelling process can be further reduced to a two-level transition problem, which can be solved analytically (see Appendix \ref{AppendixtunnellingFrequency}). 
In this case, the tunnelling $| 0_+, + \rangle \leftrightarrow | 1_-, - \rangle $ is negligible. 
On the other hand, if $\epsilon=\omega$ and levels $| 0_+, + \rangle$ and $|1_-,-\rangle$ are degenerate, the $|0_+,+\rangle \leftrightarrow | 0_-, - \rangle$ tunnelling can be omitted and the $|1_+,+\rangle \leftrightarrow | 0_-, - \rangle$ process dominates. 

In the following we consider the tunnelling dynamics with various values of $\epsilon$. 
Since our computations only concern the lowest several levels, we restrict ourselves in one cycle, i.e., $0\le \epsilon/\epsilon_c \lesssim 1$ without loss of generality.
In the intermediate regime, it is meaningless to calculate tunnelling frequencies. 
We thus use the interpolating formula
\begin{equation}\label{UnifiedFrequency}
\omega_u = (1-\epsilon)\omega_{00} + \epsilon \omega_{01},\quad T=\dfrac{2\pi}{\omega_u}, 
\end{equation}
to rescale the time axis in the dynamics computation of the AQRM.

\begin{figure}[htbp]\centering         
	
	\subfigure{                
		\centering                                                 
		\includegraphics[width=.9\linewidth]{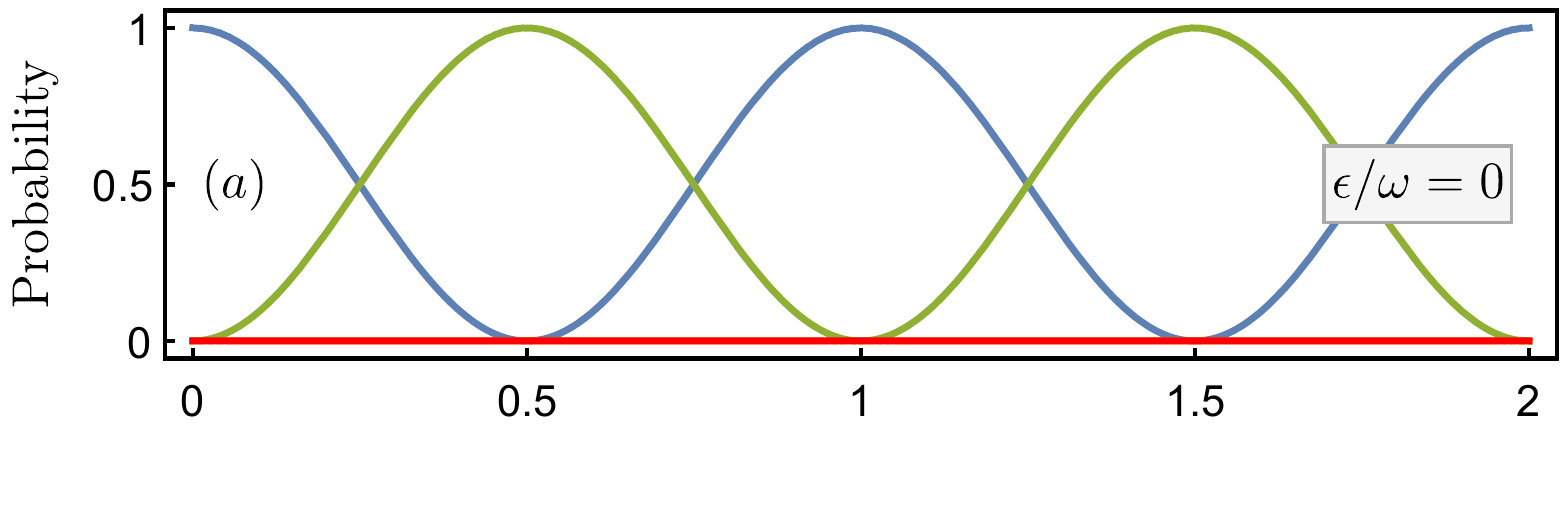}     
		\label{TunnellingAQRM1}   
	}   
	\\[-4.6ex]
	\subfigure{                
		\centering                                                 
		\includegraphics[width=.9\linewidth]{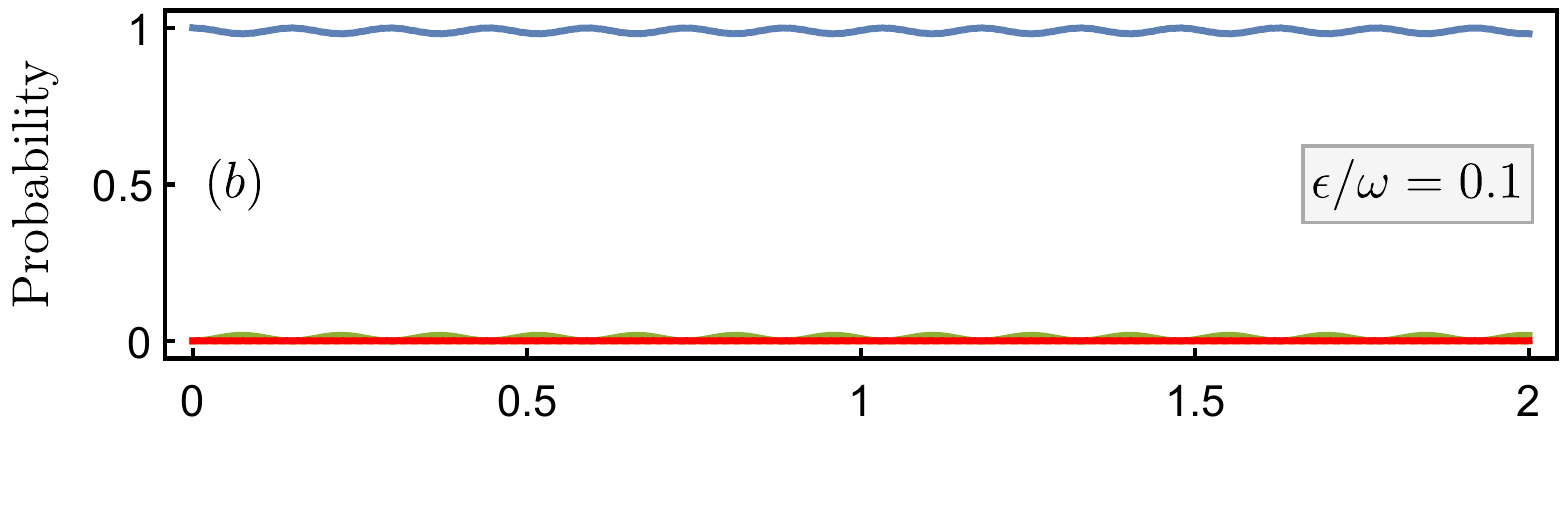}      
		\label{TunnellingAQRM2}     
	}   
	\\[-4.6ex]
	\subfigure{                
		\centering                                                 
		\includegraphics[width=.9\linewidth]{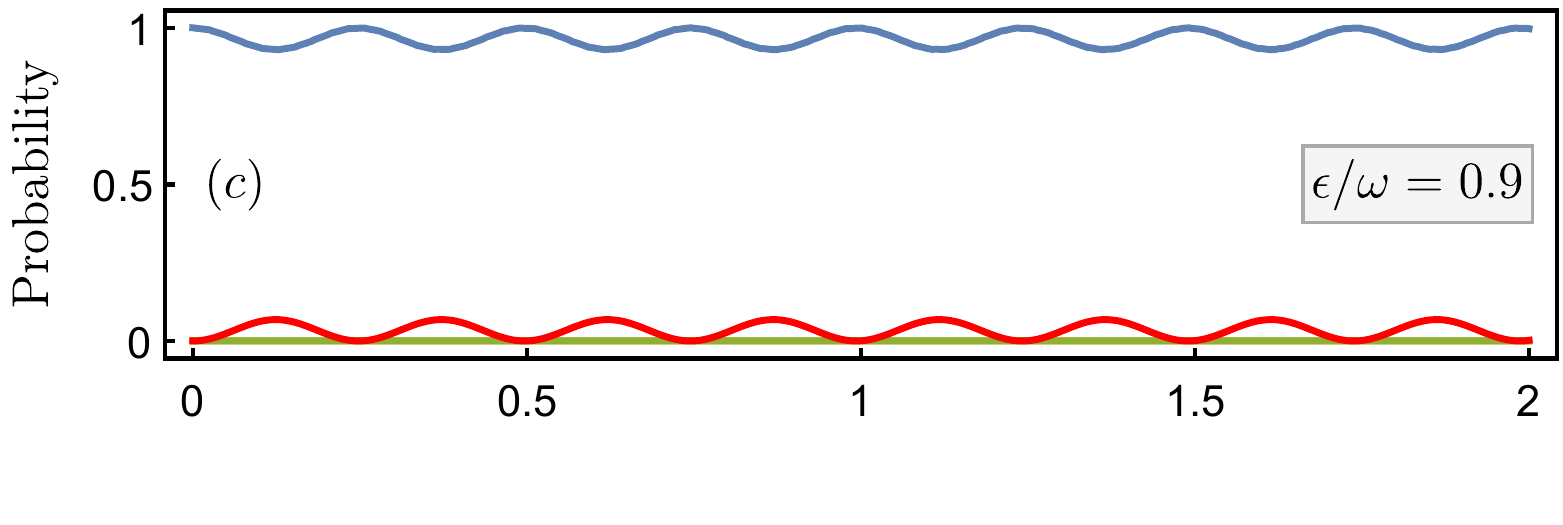}    
		\label{TunnellingAQRM3}       
	}   
	\\[-4.6ex]
	\subfigure{                
		\centering                                                 
		\includegraphics[width=.9\linewidth]{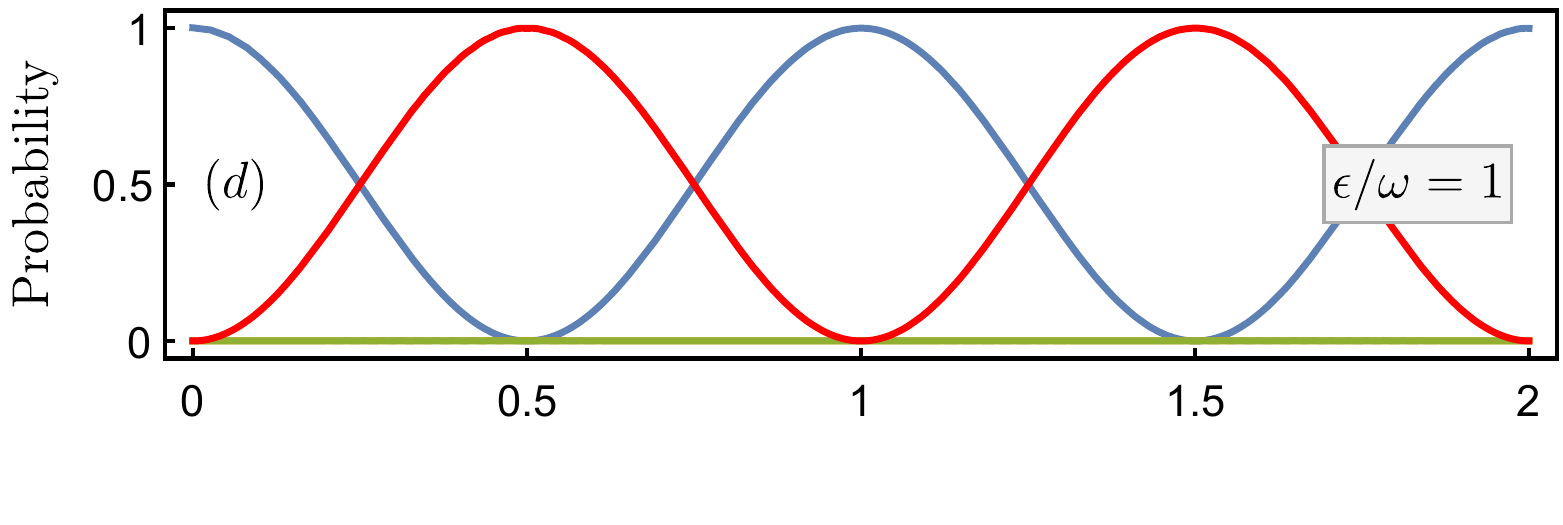}      
		\label{TunnellingAQRM4}     
	}   
	\\[-4.6ex]
	\subfigure{                
		\centering                                                 
		\includegraphics[width=.9\linewidth]{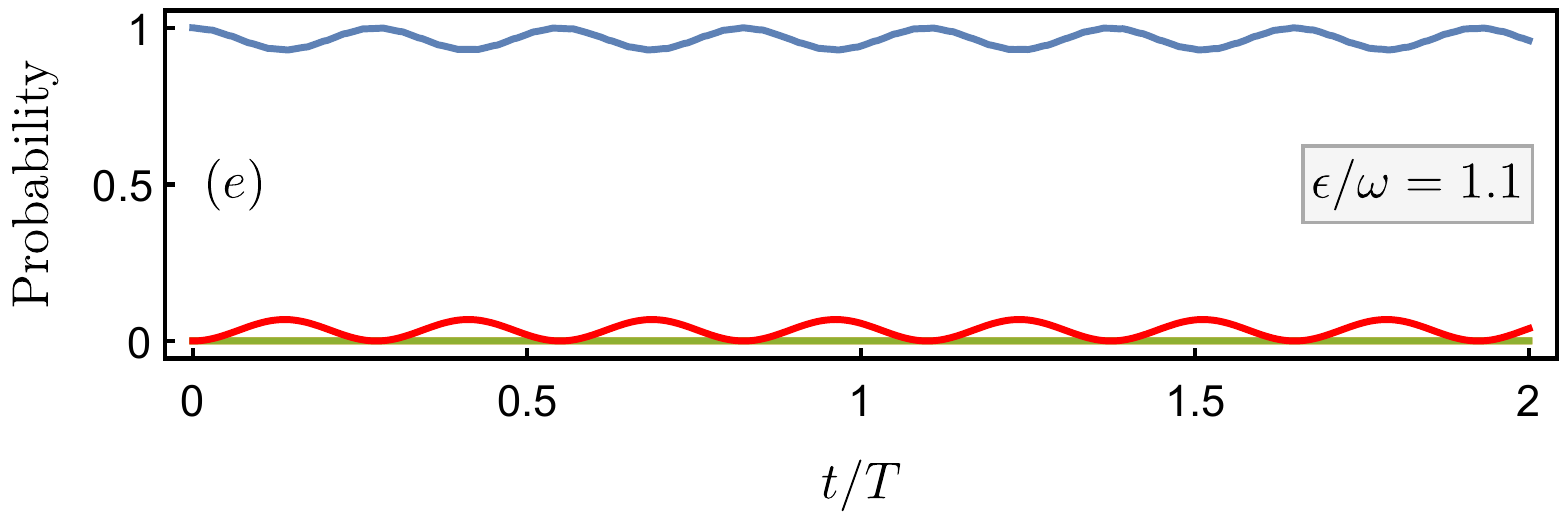}      
		\label{TunnellingAQRM5}     
	}   
	\\[-1ex]
	\subfigure{                
		\centering                                                 
		\includegraphics[width=0.72\linewidth]{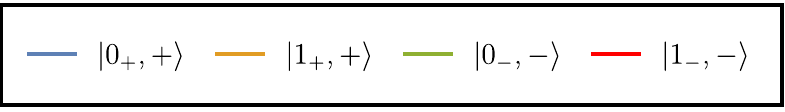}        
	}  
	\caption{Tunnelling dynamics of the asymmetric quantum Rabi model, from initial state $|0_+,+\rangle$ (blue) to $|0_-,-\rangle$ (green), $|1_+,+\rangle$ (orange) and $ |1_-,-\rangle$ (red) with various indicated values of $\epsilon$. Other parameters are $\Delta=0.1$, $g=1$ and $\omega=1$. Time is rescaled with $T$ determined by Eq.~(\ref{UnifiedFrequency}). The population of state $|1_+,+\rangle$ is almost always zero and invisible in the figures, which justifies the three-level approximation. } 
	\label{TunnellingAQRM}
\end{figure}

The time evolution of the AQRM for various values of $\epsilon$ is shown in Fig.~\ref{TunnellingAQRM}.
Here we have chosen small tunnelling parameter $\Delta/\omega=0.1$ to emphasize the peculiarity of the special cases where $\epsilon$-condition (\ref{aqrmcondition}) is satisfied. 
However, we shall see later that the phenomena shown here are rather general and apply to larger $\Delta$ cases. 
In Fig.~\ref{TunnellingAQRM}, when $\epsilon=0$, i.e., the symmetric case, the tunnelling almost only occurs between the degenerate states $|0_+,+\rangle$ and $|0_-,-\rangle$, behaving like Rabi oscillation in a two level system. 
Then we increase $\epsilon/\omega$ to 0.1, the tunnelling decreases drastically to almost vanish.
With $\epsilon/\omega=1$, the $\epsilon$-condition (\ref{aqrmcondition}) is satisfied, and the tunnelling oscillation takes place again, between now degenerate states $|0_+,+\rangle$ and $|1_-,-\rangle$. 
If we further increase $\epsilon$ to $1.1\omega$, the tunnelling decreases again. 
Thus we see that when $\epsilon/\omega=n$, the tunnelling exclusively occurs within the degenerate states. 
This is, however, not so obvious in the normal photon-qubit basis. 
It is also worth noting that the population of state $|1_+,+\rangle$ is almost always zero, which justifies the three-level approximation shown in 
Fig.~\ref{3Level}.

\begin{figure}[htbp]\centering                                              
	\subfigure{                
		\centering                                                 
		\includegraphics[width=.9\linewidth]{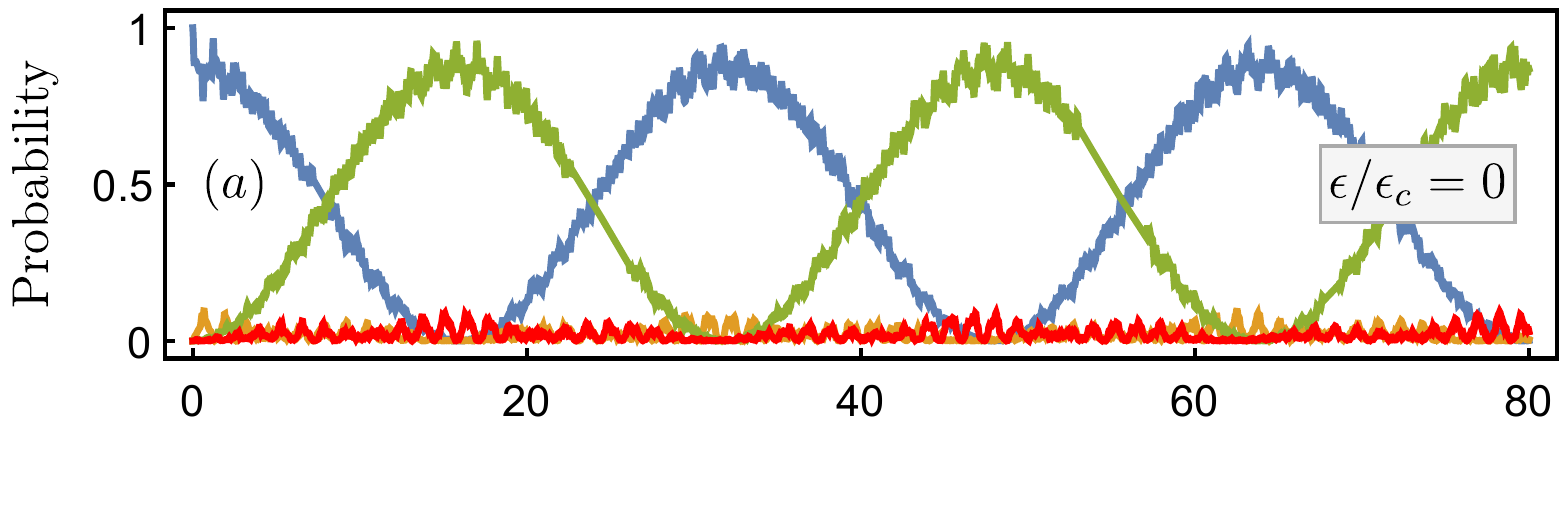}	\label{TunnellingARSM1}        
	}   
	\\[-4.5ex]
	\subfigure{                
		\centering                                                 
		\includegraphics[width=.9\linewidth]{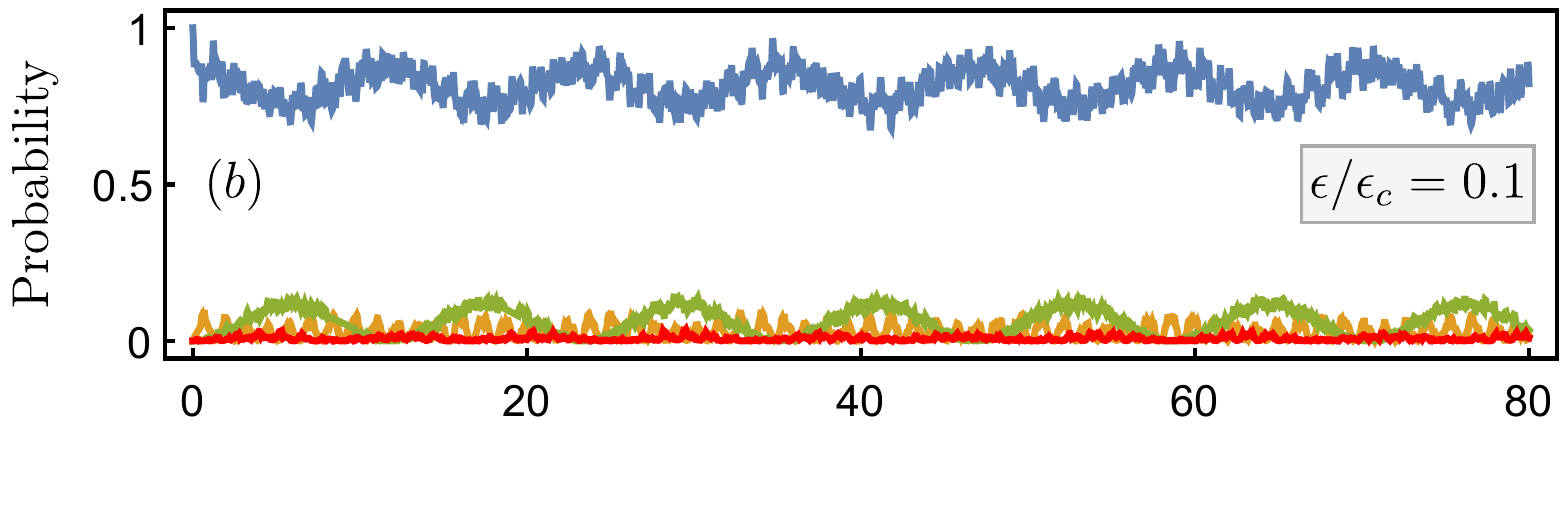}	\label{TunnellingARSM2}        
	}   
	\\[-4.5ex]
	\subfigure{                
		\centering                                                 
		\includegraphics[width=.9\linewidth]{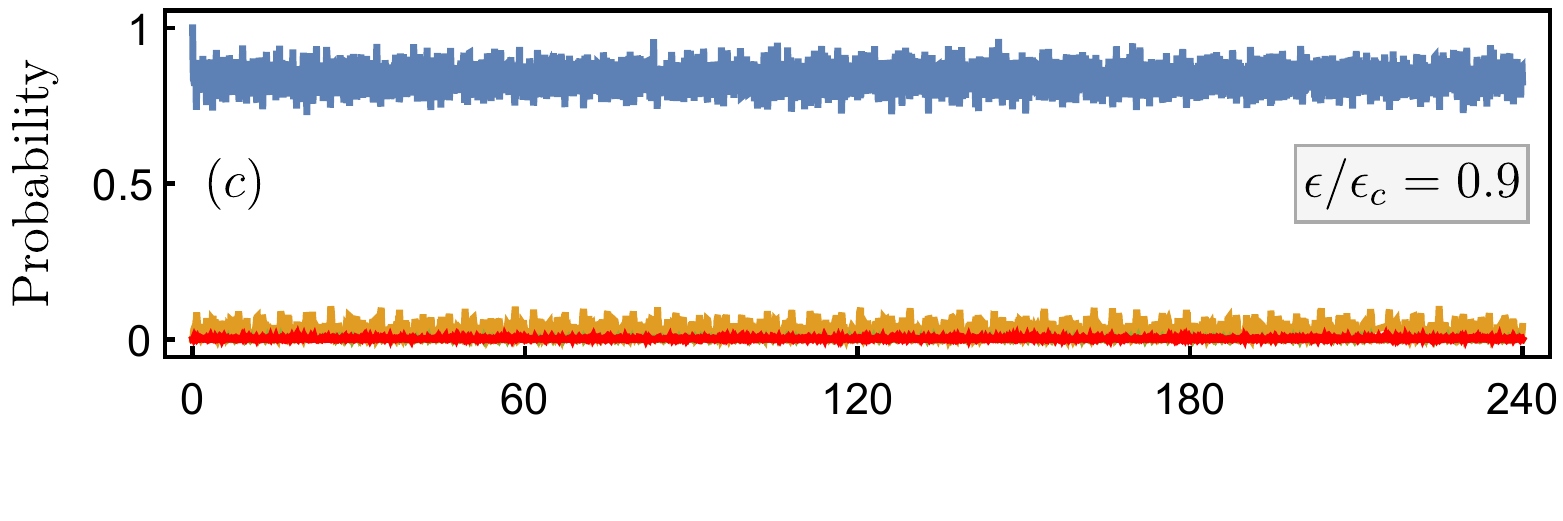}	\label{TunnellingARSM3}        
	}   
	\\[-4.5ex]
	\subfigure{                
		\centering                                                 
		\includegraphics[width=.9\linewidth]{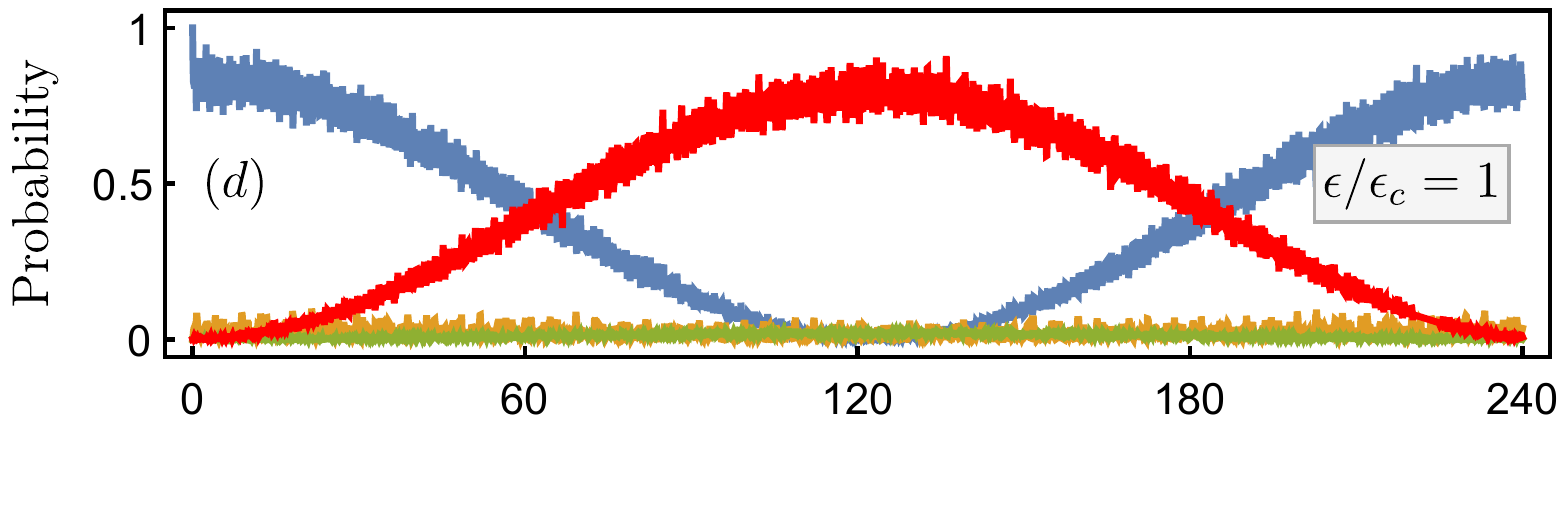}	\label{TunnellingARSM4}        
	}   
	\\[-4.5ex]
	\subfigure{                
		\centering                                                 
		\includegraphics[width=.9\linewidth]{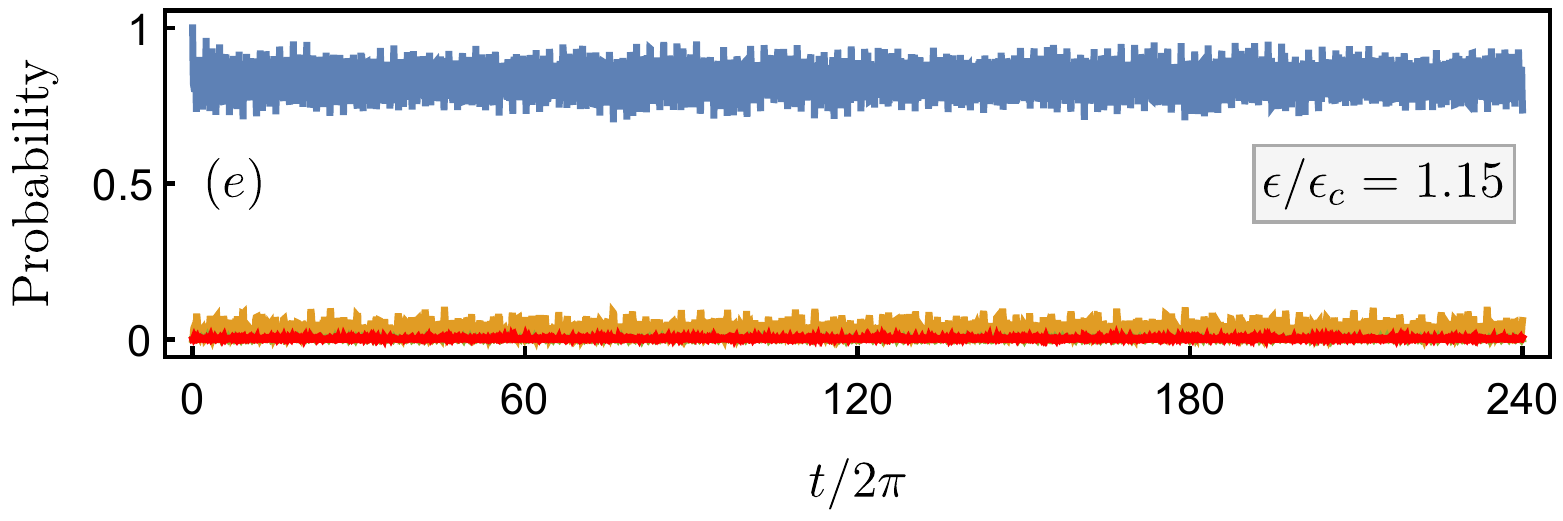}	\label{TunnellingARSM5}        
	}  
	\\[-1ex]
	\subfigure{                
		\centering                                                 
		\includegraphics[width=0.72\linewidth]{TunnellingLegends.pdf}        
	}  
	\caption{Tunnelling dynamics of asymmetric Rabi-Stark model, from initial state $|0_+,+\rangle$ (blue) to $|0_-,-\rangle$ (green), $|1_+,+\rangle$ (orange) and $ |1_-,-\rangle$ (red) with various indicated values of $\epsilon$. Other parameters are $\Delta=0.8$, $U=0.5$, $g=1$, $\omega=1$, and $\epsilon_c$ is determined through Eq.~(\ref{arsmcondition}).  } 
	\label{TunnellingARSM}
\end{figure}

\subsection{Tunnelling Dynamics in the ARSM}

The picture is much more complicated in the ARSM due to the nonlinear tunnelling term
\begin{equation}\label{key}
H^{RS}_t = \left(\dfrac{\Delta}{2} + U a^\dagger a \right) \sigma_z. 
\end{equation}
The three-level approximation displayed in Fig.~\ref{3Level} is no longer valid, because the Stark term substantially enhances the tunnelling between levels with large energy gaps. 
As a consequence, the tunnelling frequencies are not easily calculated in the ARSM. 
Fortunately, with the analytic $\epsilon$-condition for the ARSM (\ref{arsmcondition}) at hand, we are still able to investigate the tunnelling dynamics numerically. 
An important difference here is, when Eq.~(\ref{arsmcondition}) is satisfied, there are no degenerate levels in the displaced oscillators. 

We now set large tunnelling parameters $\Delta/\omega=0.8$ and $U/\omega=0.5$, which implies $\epsilon_c\approx 0.866$ according to 
Eq.~(\ref{arsmcondition}). 
As a result, $\epsilon/\epsilon_c=1.15$ corresponds to the degenerate case. 
The time evolution is shown in Fig.~\ref{TunnellingARSM}.
The overall behavior of tunnelling dynamics in the ARSM is quite similar to that of the AQRM displayed in Fig.~\ref{TunnellingAQRM}. 
Particularly, we notice that the tunnelling oscillation emerges when $\epsilon/\epsilon_c=1$, in Fig.~\ref{TunnellingARSM4}, rather than the degenerate case in Fig.~\ref{TunnellingARSM5}. 
This confirms the special role of $\epsilon$-condition (\ref{arsmcondition}) in the ARSM. 

Here we observe that large tunnelling parameters $\Delta$ and $U$ lead to similar dynamics but generate extra small-amplitude fast oscillations.

\section{Conclusion}\label{SectionConclusion}
In this work, we have explored hidden symmetry in detail for various fundamental light-matter interaction models.
We find that the hidden symmetry is not limited to the AQRM, where it was first observed, 
but is also present in other models with an asymmetric qubit bias term. 
We have determined the conditions for the existence of this hidden symmetry in 
the ARSM (\ref{harsm}) and in the anisotropic AQRM (\ref{anisotropicQRM}) by 
taking advantage of the analytic solutions for these models.
To the relatively simple known condition (\ref{aqrmcondition}) for the AQRM we thus add the condition 
(\ref{arsmcondition}) for the ARSM and the condition (\ref{anqrmcondition}) for the anisotropic AQRM. 
We conjecture that this hidden symmetry is a rather general property in the light-matter interaction models with broken $\mathbb{Z}_2$ symmetry.
It's possible existence in other related, but more complicated, models is also discussed. 
Here we have not further pursued the $\epsilon$-conditions for these models, which in principle look achievable. 

The hidden symmetry in the displaced oscillator basis and the tunnelling dynamics in the 
AQRM (\ref{haqrm}) and the ARSM (\ref{harsm}) have also been investigated.
In the symmetric cases where $\epsilon=0$, the tunnelling dominantly takes place between the degenerate levels. 
On the other hand, when $\epsilon$-conditions (\ref{aqrmcondition}) and (\ref{arsmcondition}) are satisfied by nonzero $\epsilon$, the tunnelling almost exclusively occurs between the correlated levels. 
In the intermediate regimes, however, tunnelling is drastically reduced. 
The time evolution of the AQRM and the ARSM reveals a strong connection between the hidden symmetry and selective tunnelling.

~

{\bf{Acknowledgments.}}
This work is supported by the Australian Research Council grant DP170104934 and the 
National Natural Science Foundation of China (Grant No. 11174375).

\appendix

\section{Displaced oscillator basis}\label{AppendixDisplacedOscillator}
We consider the displaced oscillator Hamiltonian
\begin{equation}\label{HdoAA}
H_\text{do} =  \omega a^\dagger a +  g\sigma_x ({a^\dagger} + a) + \dfrac{\epsilon}{2}\sigma_x . 
\end{equation}
To diagonalize Eq.~(\ref{HdoAA}), we first write the eigenstates as
\begin{equation}\label{EigenstatesAA}
|\phi_\pm , \pm \rangle = | \phi_\pm \rangle \otimes |\pm\rangle, 
\end{equation}
where $\sigma_x|\pm \rangle = \pm | \pm \rangle$ and $|\phi_\pm\rangle$ are to be determined. 
The time-independent Schr\"odinger equation for Eq.~(\ref{HdoAA}) is then
\begin{equation}\label{SEAA}
\left[\pm g(a^\dagger + a) + \omega a^\dagger a \right]|\phi_\pm\rangle = \left( E\mp \dfrac{\epsilon}{2}\right)|\phi_\pm\rangle. 
\end{equation}
The l.h.s can be written as
\begin{equation}\label{TransformationAA}
\left[\left(a^\dagger\pm \dfrac{g}{\omega}\right)\left(a \pm \dfrac{g}{\omega} \right)\right] = \mathcal{D}^\dagger\left(\pm \dfrac{g}{\omega}\right) a^\dagger a \mathcal{D}\left(\pm \dfrac{g}{\omega}\right), 
\end{equation}
where $\mathcal{D}(\alpha) = \exp[\alpha(a^\dagger- a)]$ is a displacement operator. 
In the position-momentum representation
\begin{equation}\label{CreationAnnihilationAA}
x = \sqrt{\dfrac{1}{2\omega}}(a^\dagger+a),\quad p=\mathrm{i}\sqrt{\dfrac{\omega}{2}}(a^\dagger - a), 
\end{equation}
the form of Eq.~(\ref{TransformationAA}) can then be understood as an harmonic oscillator displaced by position $\mp\frac{g}{\omega} x_0$, with $x_0 = \sqrt{1/2\omega}$. 
The eigenstates of the displaced oscillators are the displaced Fock states, also referred to as generalized coherent states, 
\begin{equation}\label{DisplacedFockAA}
|\phi_\pm\rangle = \exp\left[\mp\dfrac{g}{\omega}\left(a^\dagger - a\right)\right]|n\rangle = |n_\pm\rangle,
\end{equation}
with $n=0,1,2,3,\dots$
The corresponding energies are given by
\begin{equation}\label{EigenvaluesAA}
E_{n}^{\pm} = n \omega -\dfrac{g^2}{\omega} \pm \dfrac{\epsilon}{2}. 
\end{equation}

The tensor product eigenstates Eq.~(\ref{EigenstatesAA}) themselves are orthonormal, 
\begin{equation}\label{OrthonormalAA}
\langle m_{j},j| n_{k},k \rangle = \delta_{mn}\delta_{jk}. 
\end{equation}
However, the displaced Fock states in Eq.~(\ref{DisplacedFockAA}) are not always orthogonal. 
Indeed, we have
\begin{equation}\label{NotOrthogoanlAA}
\langle m_\pm | n_\pm \rangle = \delta_{mn},\quad \langle m_\pm | n_\mp \rangle \ne 0. 
\end{equation}
The overlap $\langle m_\pm | n_\mp \rangle \ne 0$ is introduced by the tunnelling terms containing $\sigma_z$ that we consider in the present paper. 

The overlap between Fock states displaced in different directions is calculated through $(m \ge n)$
\begin{equation}\label{OverlapAA}
\langle m_- | n_+ \rangle = e^{-2\alpha^2} (-2\alpha)^{m-n}\sqrt{\dfrac{n!}{m!}} L_N^{m-n}(4\alpha^2) ,
\end{equation}
where $\alpha = g/\omega$ and $L_k^j$ is the associated Laguerre polynomial. 
For $m<n$, we exploit the identity $\langle m_- | n_+ \rangle = (-1)^{n-m}\langle n_- | m_+ \rangle$.
Since this overlap is always real, we also know that  
\begin{equation}\label{Overlap2AA}
\langle m_- | n_+ \rangle = \langle n_+ | m_- \rangle.
\end{equation}
We can now write the Hamiltonian matrix of the AQRM or the ARSM in the displaced oscillator basis 
$\{ | m_-,-\rangle,\dots,| n_+,+ \rangle,\dots \}$ as
\begin{equation}\label{MatrixAA}
H = \begin{pmatrix}
E_{0}^{-} & 0 & 0 & \cdots & \Omega_{00} & \Omega_{01} & \Omega_{02} & \cdots \\
0 & E_{1}^{-} & 0 & \cdots & \Omega_{10} & \Omega_{11} & \Omega_{12} & \cdots \\
0 & 0 & E_{2}^{-} & \cdots & \Omega_{20} & \Omega_{21} & \Omega_{22} & \cdots \\
\vdots & \vdots & \vdots & \ddots & \vdots & \vdots & \vdots & \cdots \\
\Omega_{00} & \Omega_{10} & \Omega_{20} & \cdots & E_{0}^{+} & 0 & 0 & \cdots \\
\Omega_{01} & \Omega_{11} & \Omega_{21} & \cdots & 0 & E_{1}^{+} & 0 & \cdots \\
\Omega_{02} & \Omega_{12} & \Omega_{22} & \cdots & 0 & 0 & E_{2}^{+} & \cdots \\
\vdots & \vdots & \vdots & \vdots & \vdots & \vdots & \vdots & \ddots \\
\end{pmatrix},
\end{equation}
where $E_{n}^{\pm}$ are determined from Eq.~(\ref{EigenvaluesAA}) and $\Omega_{mn}$ denotes the matrix element $\langle m_-,-|H_t|n_ +,+\rangle$. Here $H_t$ is the tunnelling term in different models and the hermiticity of the Hamiltonian has been used. 

From now on, in the appendix A we will exclusively work in the order $\langle m_- | n_+ \rangle$, we thus omit the subscripts and simply write the nonzero overlap as $\langle m | n \rangle$. 
In the AQRM (\ref{haqrm}), the tunnelling term is
\begin{equation}\label{AQRMtunnellingTermAA}
H_t^a = \dfrac{\Delta}{2}\sigma_z,
\end{equation}
and the resulting matrix elements are thus simply
\begin{equation}\label{AQRMMatrixElemtnsAA}
\Omega_{mn}^a = \dfrac{\Delta}{2} \langle m | n \rangle. 
\end{equation}

In the ARSM (\ref{harsm}), the tunnelling term reads
\begin{equation}\label{ARSMtunnellingTermAA}
H_t^{\mathrm{RS}} = \left(\dfrac{\Delta}{2} + U a^\dagger a\right)\sigma_z,
\end{equation}
and the matrix elements are
\begin{equation}\label{ARSMMatrixElementsAA}
\Omega_{mn}^{\mathrm{RS}} = \dfrac{\Delta}{2} \langle m | n \rangle + U \langle m |a^\dagger a| n \rangle.
\end{equation}
After some commutator operations, we have
\begin{equation}\label{key}
a\mathcal{D(\alpha)} = \mathcal{D(\alpha)}(a-\alpha), 
\end{equation}
the second term in Eq.~(\ref{ARSMMatrixElementsAA}) is then obtained as
\begin{equation}\label{ARSMMatrixElements2AA}
\begin{aligned}
\langle m |a^\dagger a| n \rangle = &\alpha\sqrt{n}\langle m | n-1 \rangle  - \alpha\sqrt{m}\langle m-1 | n \rangle \\
&+\sqrt{mn}\langle m-1 | n-1 \rangle -\alpha^2\langle m | n \rangle ,
\end{aligned}
\end{equation}
with aforementioned $\alpha=g/\omega$. 

Therefore, all the matrix elements in Eq.~(\ref{MatrixAA}) for the AQRM and the ARSM can be calculated. 

~

\section{Tunnelling frequency}\label{AppendixtunnellingFrequency}
The tunnelling frequency between any two interacting levels is calculated using standard time-dependent perturbation theory \cite{J.J.Sakurai2017, DavidJ.Griffiths2018}. 
Suppose the system Eq.~(\ref{HdoAA}) is initially in the state $|m_+,+\rangle$ with energy $E_{m}^{+}$ and only interacts with state $|n_-,-\rangle$ having energy $E_{n}^{-}$.  
The off-diagonal perturbation is turned on at time $t=0$. 
According to time-dependent Schr\"odinger equation, the coefficients of two states at time $t$ obey the differential equations
\begin{equation}\label{ODEAB}
\dot{c}_m = -\mathrm{i}\Omega_{mn} e^{-\mathrm{i}\delta_{nm}t} c_n, \quad  \dot{c}_n = -\mathrm{i} \Omega_{nm} e^{\mathrm{i}\delta_{nm}t} c_m,
\end{equation}
with  $\delta_{nm} = E_{n}^{-} - E_{m}^{+}$ being the energy gap. 
Eq.~(\ref{ODEAB}) can be solved exactly as
\begin{equation}\label{CoefficientsAB}
\begin{aligned}
c_m(t) =& \exp\left(\dfrac{-\mathrm{i}\delta_{mn} t}{2}\right)\times \\
&\left[\cos\left(\omega_{mn} t\right) + \dfrac{\mathrm{i}\delta_{nm}}{\omega_{mn}}\sin\left(\omega_{mn} t\right)\right] ,
\end{aligned}
\end{equation}
and
\begin{equation}\label{Coefficients2AB}
c_n(t) = \dfrac{\Omega_{nm}}{\mathrm{i}\omega_{mn}}\exp\left(\dfrac{\mathrm{i}\delta_{nm} t}{2}\right)\sin\left(\omega_{mn} t\right),
\end{equation}
respectively, where the tunnelling frequency $\omega_{mn} = \sqrt{\delta_{nm}^2 + 4\Omega_{mn}^2}/2$. 
The corresponding populations are therefore
\begin{equation}\label{PopulationAB}
P_m(t) = \cos^2\left(\omega_{mn} t\right)+ \dfrac{\delta_{nm}^2}{4\omega_{mn}^2}\sin^2\left(\omega_{mn} t\right),
\end{equation}
and
\begin{equation}\label{Population2AB}
P_n(t) = \dfrac{\Omega_{mn}^2}{\omega_{mn}^2}\sin^2\left(\omega_{mn} t\right). 
\end{equation}

The tunnelling dynamics in Fig.~\ref{TunnellingAQRM} can be perfectly described by these last two equations. 

%\bibliography{../../Bibliography/RabiModelBib}

\begin{thebibliography}{46}%
	\makeatletter
	\providecommand \@ifxundefined [1]{%
		\@ifx{#1\undefined}
	}%
	\providecommand \@ifnum [1]{%
		\ifnum #1\expandafter \@firstoftwo
		\else \expandafter \@secondoftwo
		\fi
	}%
	\providecommand \@ifx [1]{%
		\ifx #1\expandafter \@firstoftwo
		\else \expandafter \@secondoftwo
		\fi
	}%
	\providecommand \natexlab [1]{#1}%
	\providecommand \enquote  [1]{``#1''}%
	\providecommand \bibnamefont  [1]{#1}%
	\providecommand \bibfnamefont [1]{#1}%
	\providecommand \citenamefont [1]{#1}%
	\providecommand \href@noop [0]{\@secondoftwo}%
	\providecommand \href [0]{\begingroup \@sanitize@url \@href}%
	\providecommand \@href[1]{\@@startlink{#1}\@@href}%
	\providecommand \@@href[1]{\endgroup#1\@@endlink}%
	\providecommand \@sanitize@url [0]{\catcode `\\12\catcode `\$12\catcode
		`\&12\catcode `\#12\catcode `\^12\catcode `\_12\catcode `\%12\relax}%
	\providecommand \@@startlink[1]{}%
	\providecommand \@@endlink[0]{}%
	\providecommand \url  [0]{\begingroup\@sanitize@url \@url }%
	\providecommand \@url [1]{\endgroup\@href {#1}{\urlprefix }}%
	\providecommand \urlprefix  [0]{URL }%
	\providecommand \Eprint [0]{\href }%
	\providecommand \doibase [0]{http://dx.doi.org/}%
	\providecommand \selectlanguage [0]{\@gobble}%
	\providecommand \bibinfo  [0]{\@secondoftwo}%
	\providecommand \bibfield  [0]{\@secondoftwo}%
	\providecommand \translation [1]{[#1]}%
	\providecommand \BibitemOpen [0]{}%
	\providecommand \bibitemStop [0]{}%
	\providecommand \bibitemNoStop [0]{.\EOS\space}%
	\providecommand \EOS [0]{\spacefactor3000\relax}%
	\providecommand \BibitemShut  [1]{\csname bibitem#1\endcsname}%
	\let\auto@bib@innerbib\@empty
	%</preamble>
	\bibitem [{\citenamefont {Rabi}(1936)}]{Rabi_1936}%
	\BibitemOpen
	\bibfield  {author} {\bibinfo {author} {\bibfnamefont {I.~I.}\ \bibnamefont
			{Rabi}},\ }\href {\doibase 10.1103/physrev.49.324} {\bibfield  {journal}
		{\bibinfo  {journal} {Phys. Rev.}\ }\textbf {\bibinfo {volume} {49}},\
		\bibinfo {pages} {324} (\bibinfo {year} {1936})}\BibitemShut {NoStop}%
	\bibitem [{\citenamefont {Rabi}(1937)}]{Rabi_1937}%
	\BibitemOpen
	\bibfield  {author} {\bibinfo {author} {\bibfnamefont {I.~I.}\ \bibnamefont
			{Rabi}},\ }\href {\doibase 10.1103/physrev.51.652} {\bibfield  {journal}
		{\bibinfo  {journal} {Phys. Rev.}\ }\textbf {\bibinfo {volume} {51}},\
		\bibinfo {pages} {652} (\bibinfo {year} {1937})}\BibitemShut {NoStop}%
	\bibitem [{\citenamefont {Fox}(2006)}]{Fox2006}%
	\BibitemOpen
	\bibfield  {author} {\bibinfo {author} {\bibfnamefont {M.}~\bibnamefont
			{Fox}},\ }\href
	{https://www.ebook.de/de/product/5214114/mark_fox_quantum_optics.html} {\emph
		{\bibinfo {title} {Quantum Optics}}}\ (\bibinfo  {publisher} {Oxford
		University Press},\ \bibinfo {year} {2006})\BibitemShut {NoStop}%
	\bibitem [{\citenamefont {Forn-D\'{\i}az}\ \emph {et~al.}(2010)\citenamefont
		{Forn-D\'{\i}az}, \citenamefont {Lisenfeld}, \citenamefont {Marcos},
		\citenamefont {Garc\'{\i}a-Ripoll}, \citenamefont {Solano}, \citenamefont
		{Harmans},\ and\ \citenamefont {Mooij}}]{Forn_D_az_2010}%
	\BibitemOpen
	\bibfield  {author} {\bibinfo {author} {\bibfnamefont {P.}~\bibnamefont
			{Forn-D\'{\i}az}}, \bibinfo {author} {\bibfnamefont {J.}~\bibnamefont
			{Lisenfeld}}, \bibinfo {author} {\bibfnamefont {D.}~\bibnamefont {Marcos}},
		\bibinfo {author} {\bibfnamefont {J.~J.}\ \bibnamefont {Garc\'{\i}a-Ripoll}},
		\bibinfo {author} {\bibfnamefont {E.}~\bibnamefont {Solano}}, \bibinfo
		{author} {\bibfnamefont {C.~J. P.~M.}\ \bibnamefont {Harmans}}, \ and\
		\bibinfo {author} {\bibfnamefont {J.~E.}\ \bibnamefont {Mooij}},\ }\href
	{\doibase 10.1103/PhysRevLett.105.237001} {\bibfield  {journal} {\bibinfo
			{journal} {Phys. Rev. Lett.}\ }\textbf {\bibinfo {volume} {105}},\ \bibinfo
		{pages} {237001} (\bibinfo {year} {2010})}\BibitemShut {NoStop}%
	\bibitem [{\citenamefont {Yoshihara}\ \emph {et~al.}(2018)\citenamefont
		{Yoshihara}, \citenamefont {Fuse}, \citenamefont {Ao}, \citenamefont
		{Ashhab}, \citenamefont {Kakuyanagi}, \citenamefont {Saito}, \citenamefont
		{Aoki}, \citenamefont {Koshino},\ and\ \citenamefont
		{Semba}}]{Yoshihara_2018}%
	\BibitemOpen
	\bibfield  {author} {\bibinfo {author} {\bibfnamefont {F.}~\bibnamefont
			{Yoshihara}}, \bibinfo {author} {\bibfnamefont {T.}~\bibnamefont {Fuse}},
		\bibinfo {author} {\bibfnamefont {Z.}~\bibnamefont {Ao}}, \bibinfo {author}
		{\bibfnamefont {S.}~\bibnamefont {Ashhab}}, \bibinfo {author} {\bibfnamefont
			{K.}~\bibnamefont {Kakuyanagi}}, \bibinfo {author} {\bibfnamefont
			{S.}~\bibnamefont {Saito}}, \bibinfo {author} {\bibfnamefont
			{T.}~\bibnamefont {Aoki}}, \bibinfo {author} {\bibfnamefont {K.}~\bibnamefont
			{Koshino}}, \ and\ \bibinfo {author} {\bibfnamefont {K.}~\bibnamefont
			{Semba}},\ }\href {\doibase 10.1103/PhysRevLett.120.183601} {\bibfield
		{journal} {\bibinfo  {journal} {Phys. Rev. Lett.}\ }\textbf {\bibinfo
			{volume} {120}},\ \bibinfo {pages} {183601} (\bibinfo {year}
		{2018})}\BibitemShut {NoStop}%
	\bibitem [{\citenamefont {Blais}\ \emph {et~al.}(2020)\citenamefont {Blais},
		\citenamefont {Grimsmo}, \citenamefont {Girvin},\ and\ \citenamefont
		{Wallraff}}]{Blais2020}%
	\BibitemOpen
	\bibfield  {author} {\bibinfo {author} {\bibfnamefont {A.}~\bibnamefont
			{Blais}}, \bibinfo {author} {\bibfnamefont {A.~L.}\ \bibnamefont {Grimsmo}},
		\bibinfo {author} {\bibfnamefont {S.~M.}\ \bibnamefont {Girvin}}, \ and\
		\bibinfo {author} {\bibfnamefont {A.}~\bibnamefont {Wallraff}},\ }\href@noop
	{} {\  (\bibinfo {year} {2020})},\ \Eprint {http://arxiv.org/abs/2005.12667}
	{arXiv:2005.12667 [quant-ph]} \BibitemShut {NoStop}%
	\bibitem [{\citenamefont {Xie}\ \emph {et~al.}(2017)\citenamefont {Xie},
		\citenamefont {Zhong}, \citenamefont {Batchelor},\ and\ \citenamefont
		{Lee}}]{Xie_2017}%
	\BibitemOpen
	\bibfield  {author} {\bibinfo {author} {\bibfnamefont {Q.}~\bibnamefont
			{Xie}}, \bibinfo {author} {\bibfnamefont {H.}~\bibnamefont {Zhong}}, \bibinfo
		{author} {\bibfnamefont {M.~T.}\ \bibnamefont {Batchelor}}, \ and\ \bibinfo
		{author} {\bibfnamefont {C.}~\bibnamefont {Lee}},\ }\href {\doibase
		10.1088/1751-8121/aa5a65} {\bibfield  {journal} {\bibinfo  {journal} {J.
				Phys. A: Math. Theor.}\ }\textbf {\bibinfo {volume} {50}},\ \bibinfo {pages}
		{113001} (\bibinfo {year} {2017})}\BibitemShut {NoStop}%
	\bibitem [{\citenamefont {Braak}(2019)}]{Braak_2019}%
	\BibitemOpen
	\bibfield  {author} {\bibinfo {author} {\bibfnamefont {D.}~\bibnamefont
			{Braak}},\ }\href {\doibase 10.3390/sym11101259} {\bibfield  {journal}
		{\bibinfo  {journal} {Symmetry}\ }\textbf {\bibinfo {volume} {11}},\ \bibinfo
		{pages} {1259} (\bibinfo {year} {2019})}\BibitemShut {NoStop}%
	\bibitem [{\citenamefont {Braak}(2011)}]{Braak_2011}%
	\BibitemOpen
	\bibfield  {author} {\bibinfo {author} {\bibfnamefont {D.}~\bibnamefont
			{Braak}},\ }\href {\doibase 10.1103/PhysRevLett.107.100401} {\bibfield
		{journal} {\bibinfo  {journal} {Phys. Rev. Lett.}\ }\textbf {\bibinfo
			{volume} {107}},\ \bibinfo {pages} {100401} (\bibinfo {year}
		{2011})}\BibitemShut {NoStop}%
	\bibitem [{\citenamefont {Zhong}\ \emph {et~al.}(2014)\citenamefont {Zhong},
		\citenamefont {Xie}, \citenamefont {Guan}, \citenamefont {Batchelor},
		\citenamefont {Gao},\ and\ \citenamefont {Lee}}]{Zhong_2014}%
	\BibitemOpen
	\bibfield  {author} {\bibinfo {author} {\bibfnamefont {H.}~\bibnamefont
			{Zhong}}, \bibinfo {author} {\bibfnamefont {Q.}~\bibnamefont {Xie}}, \bibinfo
		{author} {\bibfnamefont {X.}~\bibnamefont {Guan}}, \bibinfo {author}
		{\bibfnamefont {M.~T.}\ \bibnamefont {Batchelor}}, \bibinfo {author}
		{\bibfnamefont {K.}~\bibnamefont {Gao}}, \ and\ \bibinfo {author}
		{\bibfnamefont {C.}~\bibnamefont {Lee}},\ }\href {\doibase
		10.1088/1751-8113/47/4/045301} {\bibfield  {journal} {\bibinfo  {journal} {J.
				Phys. A: Math. Theor.}\ }\textbf {\bibinfo {volume} {47}},\ \bibinfo {pages}
		{045301} (\bibinfo {year} {2014})}\BibitemShut {NoStop}%
	\bibitem [{\citenamefont {Li}\ and\ \citenamefont {Batchelor}(2015)}]{Li_2015}%
	\BibitemOpen
	\bibfield  {author} {\bibinfo {author} {\bibfnamefont {Z.-M.}\ \bibnamefont
			{Li}}\ and\ \bibinfo {author} {\bibfnamefont {M.~T.}\ \bibnamefont
			{Batchelor}},\ }\href {\doibase 10.1088/1751-8113/48/45/454005} {\bibfield
		{journal} {\bibinfo  {journal} {J. Phys. A: Math. Theor.}\ }\textbf {\bibinfo
			{volume} {48}},\ \bibinfo {pages} {454005} (\bibinfo {year}
		{2015})}\BibitemShut {NoStop}%
	\bibitem [{\citenamefont {Li}\ and\ \citenamefont {Batchelor}(2016)}]{Li_2016}%
	\BibitemOpen
	\bibfield  {author} {\bibinfo {author} {\bibfnamefont {Z.-M.}\ \bibnamefont
			{Li}}\ and\ \bibinfo {author} {\bibfnamefont {M.~T.}\ \bibnamefont
			{Batchelor}},\ }\href {\doibase 10.1088/1751-8113/49/36/369401} {\bibfield
		{journal} {\bibinfo  {journal} {J. Phys. A: Math. Theor.}\ }\textbf {\bibinfo
			{volume} {49}},\ \bibinfo {pages} {369401} (\bibinfo {year}
		{2016})}\BibitemShut {NoStop}%
	\bibitem [{\citenamefont {Batchelor}\ \emph {et~al.}(2015)\citenamefont
		{Batchelor}, \citenamefont {Li},\ and\ \citenamefont {Zhou}}]{Batchelor2015}%
	\BibitemOpen
	\bibfield  {author} {\bibinfo {author} {\bibfnamefont {M.~T.}\ \bibnamefont
			{Batchelor}}, \bibinfo {author} {\bibfnamefont {Z.-M.}\ \bibnamefont {Li}}, \
		and\ \bibinfo {author} {\bibfnamefont {H.-Q.}\ \bibnamefont {Zhou}},\ }\href
	{\doibase 10.1088/1751-8113/49/1/01lt01} {\bibfield  {journal} {\bibinfo
			{journal} {J. Phys. A: Math. Theor.}\ }\textbf {\bibinfo {volume} {49}},\
		\bibinfo {pages} {01LT01} (\bibinfo {year} {2015})}\BibitemShut {NoStop}%
	\bibitem [{\citenamefont {Liu}\ \emph {et~al.}(2017)\citenamefont {Liu},
		\citenamefont {Ying}, \citenamefont {An}, \citenamefont {Luo},\ and\
		\citenamefont {Lin}}]{Liu2017a}%
	\BibitemOpen
	\bibfield  {author} {\bibinfo {author} {\bibfnamefont {M.}~\bibnamefont
			{Liu}}, \bibinfo {author} {\bibfnamefont {Z.-J.}\ \bibnamefont {Ying}},
		\bibinfo {author} {\bibfnamefont {J.-H.}\ \bibnamefont {An}}, \bibinfo
		{author} {\bibfnamefont {H.-G.}\ \bibnamefont {Luo}}, \ and\ \bibinfo
		{author} {\bibfnamefont {H.-Q.}\ \bibnamefont {Lin}},\ }\href {\doibase
		10.1088/1751-8121/aa56f6} {\bibfield  {journal} {\bibinfo  {journal} {J.
				Phys. A: Math. Theor.}\ }\textbf {\bibinfo {volume} {50}},\ \bibinfo {pages}
		{084003} (\bibinfo {year} {2017})}\BibitemShut {NoStop}%
	\bibitem [{\citenamefont {Mao}\ \emph {et~al.}(2018)\citenamefont {Mao},
		\citenamefont {Liu}, \citenamefont {Wu}, \citenamefont {Li}, \citenamefont
		{Ying},\ and\ \citenamefont {Luo}}]{Mao_2018}%
	\BibitemOpen
	\bibfield  {author} {\bibinfo {author} {\bibfnamefont {B.-B.}\ \bibnamefont
			{Mao}}, \bibinfo {author} {\bibfnamefont {M.}~\bibnamefont {Liu}}, \bibinfo
		{author} {\bibfnamefont {W.}~\bibnamefont {Wu}}, \bibinfo {author}
		{\bibfnamefont {L.}~\bibnamefont {Li}}, \bibinfo {author} {\bibfnamefont
			{Z.-J.}\ \bibnamefont {Ying}}, \ and\ \bibinfo {author} {\bibfnamefont
			{H.-G.}\ \bibnamefont {Luo}},\ }\href {\doibase
		10.1088/1674-1056/27/5/054219} {\bibfield  {journal} {\bibinfo  {journal}
			{Chin. Phys. B}\ }\textbf {\bibinfo {volume} {27}},\ \bibinfo {pages}
		{054219} (\bibinfo {year} {2018})}\BibitemShut {NoStop}%
	\bibitem [{\citenamefont {Guan}\ \emph {et~al.}(2018)\citenamefont {Guan},
		\citenamefont {Li}, \citenamefont {Dunning},\ and\ \citenamefont
		{Batchelor}}]{Guan_2018}%
	\BibitemOpen
	\bibfield  {author} {\bibinfo {author} {\bibfnamefont {K.-L.}\ \bibnamefont
			{Guan}}, \bibinfo {author} {\bibfnamefont {Z.-M.}\ \bibnamefont {Li}},
		\bibinfo {author} {\bibfnamefont {C.}~\bibnamefont {Dunning}}, \ and\
		\bibinfo {author} {\bibfnamefont {M.~T.}\ \bibnamefont {Batchelor}},\ }\href
	{\doibase 10.1088/1751-8121/aacb44} {\bibfield  {journal} {\bibinfo
			{journal} {J. Phys. A: Math. Theor.}\ }\textbf {\bibinfo {volume} {51}},\
		\bibinfo {pages} {315204} (\bibinfo {year} {2018})}\BibitemShut {NoStop}%
	\bibitem [{\citenamefont {Xie}\ \emph {et~al.}(2020{\natexlab{a}})\citenamefont
		{Xie}, \citenamefont {Mao}, \citenamefont {Li}, \citenamefont {Wang},
		\citenamefont {Sun}, \citenamefont {Wang}, \citenamefont {You},\ and\
		\citenamefont {Liu}}]{Xie2020}%
	\BibitemOpen
	\bibfield  {author} {\bibinfo {author} {\bibfnamefont {W.}~\bibnamefont
			{Xie}}, \bibinfo {author} {\bibfnamefont {B.-B.}\ \bibnamefont {Mao}},
		\bibinfo {author} {\bibfnamefont {G.}~\bibnamefont {Li}}, \bibinfo {author}
		{\bibfnamefont {W.}~\bibnamefont {Wang}}, \bibinfo {author} {\bibfnamefont
			{C.}~\bibnamefont {Sun}}, \bibinfo {author} {\bibfnamefont {Y.}~\bibnamefont
			{Wang}}, \bibinfo {author} {\bibfnamefont {W.-L.}\ \bibnamefont {You}}, \
		and\ \bibinfo {author} {\bibfnamefont {M.}~\bibnamefont {Liu}},\ }\href
	{\doibase 10.1088/1751-8121/ab4b7a} {\bibfield  {journal} {\bibinfo
			{journal} {J. Phys. A: Math. Theor.}\ }\textbf {\bibinfo {volume} {53}},\
		\bibinfo {pages} {095302} (\bibinfo {year} {2020}{\natexlab{a}})}\BibitemShut
	{NoStop}%
	\bibitem [{\citenamefont {Wakayama}(2017)}]{Wakayama_2017}%
	\BibitemOpen
	\bibfield  {author} {\bibinfo {author} {\bibfnamefont {M.}~\bibnamefont
			{Wakayama}},\ }\href {\doibase 10.1088/1751-8121/aa649b} {\bibfield
		{journal} {\bibinfo  {journal} {J. Phys. A: Math. Theor.}\ }\textbf {\bibinfo
			{volume} {50}},\ \bibinfo {pages} {174001} (\bibinfo {year}
		{2017})}\BibitemShut {NoStop}%
	\bibitem [{\citenamefont {Kimoto}\ \emph {et~al.}(2020)\citenamefont {Kimoto},
		\citenamefont {Reyes-Bustos},\ and\ \citenamefont {Wakayama}}]{Kimoto_2020}%
	\BibitemOpen
	\bibfield  {author} {\bibinfo {author} {\bibfnamefont {K.}~\bibnamefont
			{Kimoto}}, \bibinfo {author} {\bibfnamefont {C.}~\bibnamefont
			{Reyes-Bustos}}, \ and\ \bibinfo {author} {\bibfnamefont {M.}~\bibnamefont
			{Wakayama}},\ }\href {\doibase 10.1093/imrn/rnaa034} {\bibfield  {journal}
		{\bibinfo  {journal} {Int. Math. Res. Not.}\ } (\bibinfo {year} {2020}),\
		10.1093/imrn/rnaa034}\BibitemShut {NoStop}%
	\bibitem [{\citenamefont {Ashhab}(2020)}]{Ashhab_2020}%
	\BibitemOpen
	\bibfield  {author} {\bibinfo {author} {\bibfnamefont {S.}~\bibnamefont
			{Ashhab}},\ }\href {\doibase 10.1103/PhysRevA.101.023808} {\bibfield
		{journal} {\bibinfo  {journal} {Phys. Rev. A}\ }\textbf {\bibinfo {volume}
			{101}},\ \bibinfo {pages} {023808} (\bibinfo {year} {2020})}\BibitemShut
	{NoStop}%
	\bibitem [{\citenamefont {Gardas}\ and\ \citenamefont
		{Dajka}(2013)}]{Gardas_2013}%
	\BibitemOpen
	\bibfield  {author} {\bibinfo {author} {\bibfnamefont {B.}~\bibnamefont
			{Gardas}}\ and\ \bibinfo {author} {\bibfnamefont {J.}~\bibnamefont {Dajka}},\
	}\href {\doibase 10.1088/1751-8113/46/26/265302} {\bibfield  {journal}
		{\bibinfo  {journal} {J. Phys. A: Math. Theor.}\ }\textbf {\bibinfo {volume}
			{46}},\ \bibinfo {pages} {265302} (\bibinfo {year} {2013})}\BibitemShut
	{NoStop}%
	\bibitem [{\citenamefont {Judd}(1979)}]{Judd_1979}%
	\BibitemOpen
	\bibfield  {author} {\bibinfo {author} {\bibfnamefont {B.~R.}\ \bibnamefont
			{Judd}},\ }\href {\doibase 10.1088/0022-3719/12/9/010} {\bibfield  {journal}
		{\bibinfo  {journal} {J. Phys. C: Solid State Phys.}\ }\textbf {\bibinfo
			{volume} {12}},\ \bibinfo {pages} {1685} (\bibinfo {year}
		{1979})}\BibitemShut {NoStop}%
	\bibitem [{\citenamefont {Ku{\'{s}}}(1985)}]{Kus1985}%
	\BibitemOpen
	\bibfield  {author} {\bibinfo {author} {\bibfnamefont {M.}~\bibnamefont
			{Ku{\'{s}}}},\ }\href {\doibase 10.1063/1.526703} {\bibfield  {journal}
		{\bibinfo  {journal} {J. Math. Phys.}\ }\textbf {\bibinfo {volume} {26}},\
		\bibinfo {pages} {2792} (\bibinfo {year} {1985})}\BibitemShut {NoStop}%
	\bibitem [{\citenamefont {Chen}\ \emph {et~al.}(2012)\citenamefont {Chen},
		\citenamefont {Wang}, \citenamefont {He}, \citenamefont {Liu},\ and\
		\citenamefont {Wang}}]{Chen2012}%
	\BibitemOpen
	\bibfield  {author} {\bibinfo {author} {\bibfnamefont {Q.-H.}\ \bibnamefont
			{Chen}}, \bibinfo {author} {\bibfnamefont {C.}~\bibnamefont {Wang}}, \bibinfo
		{author} {\bibfnamefont {S.}~\bibnamefont {He}}, \bibinfo {author}
		{\bibfnamefont {T.}~\bibnamefont {Liu}}, \ and\ \bibinfo {author}
		{\bibfnamefont {K.-L.}\ \bibnamefont {Wang}},\ }\href {\doibase
		10.1103/PhysRevA.86.023822} {\bibfield  {journal} {\bibinfo  {journal} {Phys.
				Rev. A}\ }\textbf {\bibinfo {volume} {86}},\ \bibinfo {pages} {023822}
		(\bibinfo {year} {2012})}\BibitemShut {NoStop}%
	\bibitem [{\citenamefont {Zhong}\ \emph {et~al.}(2013)\citenamefont {Zhong},
		\citenamefont {Xie}, \citenamefont {Batchelor},\ and\ \citenamefont
		{Lee}}]{Zhong_2013}%
	\BibitemOpen
	\bibfield  {author} {\bibinfo {author} {\bibfnamefont {H.}~\bibnamefont
			{Zhong}}, \bibinfo {author} {\bibfnamefont {Q.}~\bibnamefont {Xie}}, \bibinfo
		{author} {\bibfnamefont {M.~T.}\ \bibnamefont {Batchelor}}, \ and\ \bibinfo
		{author} {\bibfnamefont {C.}~\bibnamefont {Lee}},\ }\href {\doibase
		10.1088/1751-8113/46/41/415302} {\bibfield  {journal} {\bibinfo  {journal}
			{J. Phys. A: Math. Theor.}\ }\textbf {\bibinfo {volume} {46}},\ \bibinfo
		{pages} {415302} (\bibinfo {year} {2013})}\BibitemShut {NoStop}%
	\bibitem [{\citenamefont {Maciejewski}\ \emph {et~al.}(2014)\citenamefont
		{Maciejewski}, \citenamefont {Przybylska},\ and\ \citenamefont
		{Stachowiak}}]{Maciejewski_2014}%
	\BibitemOpen
	\bibfield  {author} {\bibinfo {author} {\bibfnamefont {A.~J.}\ \bibnamefont
			{Maciejewski}}, \bibinfo {author} {\bibfnamefont {M.}~\bibnamefont
			{Przybylska}}, \ and\ \bibinfo {author} {\bibfnamefont {T.}~\bibnamefont
			{Stachowiak}},\ }\href {\doibase 10.1016/j.physleta.2014.10.001} {\bibfield
		{journal} {\bibinfo  {journal} {Phys. Lett. A}\ }\textbf {\bibinfo {volume}
			{378}},\ \bibinfo {pages} {3445} (\bibinfo {year} {2014})}\BibitemShut
	{NoStop}%
	\bibitem [{\citenamefont {Grimsmo}\ and\ \citenamefont
		{Parkins}(2013)}]{Grimsmo_2013}%
	\BibitemOpen
	\bibfield  {author} {\bibinfo {author} {\bibfnamefont {A.~L.}\ \bibnamefont
			{Grimsmo}}\ and\ \bibinfo {author} {\bibfnamefont {S.}~\bibnamefont
			{Parkins}},\ }\href {\doibase 10.1103/PhysRevA.87.033814} {\bibfield
		{journal} {\bibinfo  {journal} {Phys. Rev. A}\ }\textbf {\bibinfo {volume}
			{87}},\ \bibinfo {pages} {033814} (\bibinfo {year} {2013})}\BibitemShut
	{NoStop}%
	\bibitem [{\citenamefont {Grimsmo}\ and\ \citenamefont
		{Parkins}(2014)}]{Grimsmo_2014}%
	\BibitemOpen
	\bibfield  {author} {\bibinfo {author} {\bibfnamefont {A.~L.}\ \bibnamefont
			{Grimsmo}}\ and\ \bibinfo {author} {\bibfnamefont {S.}~\bibnamefont
			{Parkins}},\ }\href {\doibase 10.1103/PhysRevA.89.033802} {\bibfield
		{journal} {\bibinfo  {journal} {Phys. Rev. A}\ }\textbf {\bibinfo {volume}
			{89}},\ \bibinfo {pages} {033802} (\bibinfo {year} {2014})}\BibitemShut
	{NoStop}%
	\bibitem [{\citenamefont {Eckle}\ and\ \citenamefont
		{Johannesson}(2017)}]{Eckle_2017}%
	\BibitemOpen
	\bibfield  {author} {\bibinfo {author} {\bibfnamefont {H.-P.}\ \bibnamefont
			{Eckle}}\ and\ \bibinfo {author} {\bibfnamefont {H.}~\bibnamefont
			{Johannesson}},\ }\href {\doibase 10.1088/1751-8121/aa785a} {\bibfield
		{journal} {\bibinfo  {journal} {J. Phys. A: Math. Theor.}\ }\textbf {\bibinfo
			{volume} {50}},\ \bibinfo {pages} {294004} (\bibinfo {year}
		{2017})}\BibitemShut {NoStop}%
	\bibitem [{\citenamefont {Xie}\ \emph {et~al.}(2019)\citenamefont {Xie},
		\citenamefont {Duan},\ and\ \citenamefont {Chen}}]{Xie_2019}%
	\BibitemOpen
	\bibfield  {author} {\bibinfo {author} {\bibfnamefont {Y.-F.}\ \bibnamefont
			{Xie}}, \bibinfo {author} {\bibfnamefont {L.}~\bibnamefont {Duan}}, \ and\
		\bibinfo {author} {\bibfnamefont {Q.-H.}\ \bibnamefont {Chen}},\ }\href
	{\doibase 10.1088/1751-8121/ab1cf6} {\bibfield  {journal} {\bibinfo
			{journal} {J. Phys. A: Math. Theor.}\ }\textbf {\bibinfo {volume} {52}},\
		\bibinfo {pages} {245304} (\bibinfo {year} {2019})}\BibitemShut {NoStop}%
	\bibitem [{\citenamefont {Chen}\ \emph {et~al.}(2020)\citenamefont {Chen},
		\citenamefont {Xie},\ and\ \citenamefont {Chen}}]{Chen2020}%
	\BibitemOpen
	\bibfield  {author} {\bibinfo {author} {\bibfnamefont {X.-Y.}\ \bibnamefont
			{Chen}}, \bibinfo {author} {\bibfnamefont {Y.-F.}\ \bibnamefont {Xie}}, \
		and\ \bibinfo {author} {\bibfnamefont {Q.-H.}\ \bibnamefont {Chen}},\
	}\href@noop {} {\  (\bibinfo {year} {2020})},\ \Eprint
	{http://arxiv.org/abs/2001.04356} {arXiv:2001.04356 [quant-ph]} \BibitemShut
	{NoStop}%
	\bibitem [{\citenamefont {Xie}\ \emph {et~al.}(2020{\natexlab{b}})\citenamefont
		{Xie}, \citenamefont {Chen}, \citenamefont {Dong},\ and\ \citenamefont
		{Chen}}]{Xie_2020}%
	\BibitemOpen
	\bibfield  {author} {\bibinfo {author} {\bibfnamefont {Y.-F.}\ \bibnamefont
			{Xie}}, \bibinfo {author} {\bibfnamefont {X.-Y.}\ \bibnamefont {Chen}},
		\bibinfo {author} {\bibfnamefont {X.-F.}\ \bibnamefont {Dong}}, \ and\
		\bibinfo {author} {\bibfnamefont {Q.-H.}\ \bibnamefont {Chen}},\ }\href
	{\doibase 10.1103/PhysRevA.101.053803} {\bibfield  {journal} {\bibinfo
			{journal} {Phys. Rev. A}\ }\textbf {\bibinfo {volume} {101}},\ \bibinfo
		{pages} {053803} (\bibinfo {year} {2020}{\natexlab{b}})}\BibitemShut
	{NoStop}%
	\bibitem [{\citenamefont {Cong}\ \emph {et~al.}(2020)\citenamefont {Cong},
		\citenamefont {Felicetti}, \citenamefont {Casanova}, \citenamefont {Lamata},
		\citenamefont {Solano},\ and\ \citenamefont {Arrazola}}]{Cong_2020}%
	\BibitemOpen
	\bibfield  {author} {\bibinfo {author} {\bibfnamefont {L.}~\bibnamefont
			{Cong}}, \bibinfo {author} {\bibfnamefont {S.}~\bibnamefont {Felicetti}},
		\bibinfo {author} {\bibfnamefont {J.}~\bibnamefont {Casanova}}, \bibinfo
		{author} {\bibfnamefont {L.}~\bibnamefont {Lamata}}, \bibinfo {author}
		{\bibfnamefont {E.}~\bibnamefont {Solano}}, \ and\ \bibinfo {author}
		{\bibfnamefont {I.}~\bibnamefont {Arrazola}},\ }\href {\doibase
		10.1103/PhysRevA.101.032350} {\bibfield  {journal} {\bibinfo  {journal}
			{Phys. Rev. A}\ }\textbf {\bibinfo {volume} {101}},\ \bibinfo {pages}
		{032350} (\bibinfo {year} {2020})}\BibitemShut {NoStop}%
	\bibitem [{\citenamefont {Tomka}\ \emph {et~al.}(2014)\citenamefont {Tomka},
		\citenamefont {El~Araby}, \citenamefont {Pletyukhov},\ and\ \citenamefont
		{Gritsev}}]{Tomka_2014}%
	\BibitemOpen
	\bibfield  {author} {\bibinfo {author} {\bibfnamefont {M.}~\bibnamefont
			{Tomka}}, \bibinfo {author} {\bibfnamefont {O.}~\bibnamefont {El~Araby}},
		\bibinfo {author} {\bibfnamefont {M.}~\bibnamefont {Pletyukhov}}, \ and\
		\bibinfo {author} {\bibfnamefont {V.}~\bibnamefont {Gritsev}},\ }\href
	{\doibase 10.1103/PhysRevA.90.063839} {\bibfield  {journal} {\bibinfo
			{journal} {Phys. Rev. A}\ }\textbf {\bibinfo {volume} {90}},\ \bibinfo
		{pages} {063839} (\bibinfo {year} {2014})}\BibitemShut {NoStop}%
	\bibitem [{\citenamefont {Xie}\ \emph {et~al.}(2014)\citenamefont {Xie},
		\citenamefont {Cui}, \citenamefont {Cao}, \citenamefont {Amico},\ and\
		\citenamefont {Fan}}]{Xie_2014}%
	\BibitemOpen
	\bibfield  {author} {\bibinfo {author} {\bibfnamefont {Q.-T.}\ \bibnamefont
			{Xie}}, \bibinfo {author} {\bibfnamefont {S.}~\bibnamefont {Cui}}, \bibinfo
		{author} {\bibfnamefont {J.-P.}\ \bibnamefont {Cao}}, \bibinfo {author}
		{\bibfnamefont {L.}~\bibnamefont {Amico}}, \ and\ \bibinfo {author}
		{\bibfnamefont {H.}~\bibnamefont {Fan}},\ }\href {\doibase
		10.1103/PhysRevX.4.021046} {\bibfield  {journal} {\bibinfo  {journal} {Phys.
				Rev. X}\ }\textbf {\bibinfo {volume} {4}},\ \bibinfo {pages} {021046}
		(\bibinfo {year} {2014})}\BibitemShut {NoStop}%
	\bibitem [{\citenamefont {Jaynes}\ and\ \citenamefont
		{Cummings}(1963)}]{Jaynes_1963}%
	\BibitemOpen
	\bibfield  {author} {\bibinfo {author} {\bibfnamefont {E.}~\bibnamefont
			{Jaynes}}\ and\ \bibinfo {author} {\bibfnamefont {F.}~\bibnamefont
			{Cummings}},\ }\href {\doibase 10.1109/proc.1963.1664} {\bibfield  {journal}
		{\bibinfo  {journal} {Proc. {IEEE}}\ }\textbf {\bibinfo {volume} {51}},\
		\bibinfo {pages} {89} (\bibinfo {year} {1963})}\BibitemShut {NoStop}%
	\bibitem [{\citenamefont {Chilingaryan}\ and\ \citenamefont
		{Rodr{\'{\i}}guez-Lara}(2015)}]{Chilingaryan_2015}%
	\BibitemOpen
	\bibfield  {author} {\bibinfo {author} {\bibfnamefont {S.~A.}\ \bibnamefont
			{Chilingaryan}}\ and\ \bibinfo {author} {\bibfnamefont {B.~M.}\ \bibnamefont
			{Rodr{\'{\i}}guez-Lara}},\ }\href {\doibase 10.1088/0953-4075/48/24/245501}
	{\bibfield  {journal} {\bibinfo  {journal} {J. Phys. B: At., Mol. Opt.
				Phys.}\ }\textbf {\bibinfo {volume} {48}},\ \bibinfo {pages} {245501}
		(\bibinfo {year} {2015})}\BibitemShut {NoStop}%
	\bibitem [{\citenamefont {Zhang}(2013)}]{Zhang2013}%
	\BibitemOpen
	\bibfield  {author} {\bibinfo {author} {\bibfnamefont {Y.-Z.}\ \bibnamefont
			{Zhang}},\ }\href {\doibase 10.1063/1.4826356} {\bibfield  {journal}
		{\bibinfo  {journal} {J. Math. Phys.}\ }\textbf {\bibinfo {volume} {54}},\
		\bibinfo {pages} {102104} (\bibinfo {year} {2013})}\BibitemShut {NoStop}%
	\bibitem [{\citenamefont {Duan}\ \emph {et~al.}(2015)\citenamefont {Duan},
		\citenamefont {He}, \citenamefont {Braak},\ and\ \citenamefont
		{Chen}}]{Duan_2015}%
	\BibitemOpen
	\bibfield  {author} {\bibinfo {author} {\bibfnamefont {L.}~\bibnamefont
			{Duan}}, \bibinfo {author} {\bibfnamefont {S.}~\bibnamefont {He}}, \bibinfo
		{author} {\bibfnamefont {D.}~\bibnamefont {Braak}}, \ and\ \bibinfo {author}
		{\bibfnamefont {Q.-H.}\ \bibnamefont {Chen}},\ }\href {\doibase
		10.1209/0295-5075/112/34003} {\bibfield  {journal} {\bibinfo  {journal}
			{{EPL} (Europhysics Letters)}\ }\textbf {\bibinfo {volume} {112}},\ \bibinfo
		{pages} {34003} (\bibinfo {year} {2015})}\BibitemShut {NoStop}%
	\bibitem [{\citenamefont {Chilingaryan}\ and\ \citenamefont
		{Rodr{\'{\i}}guez-Lara}(2013)}]{Chilingaryan_2013}%
	\BibitemOpen
	\bibfield  {author} {\bibinfo {author} {\bibfnamefont {S.~A.}\ \bibnamefont
			{Chilingaryan}}\ and\ \bibinfo {author} {\bibfnamefont {B.~M.}\ \bibnamefont
			{Rodr{\'{\i}}guez-Lara}},\ }\href {\doibase 10.1088/1751-8113/46/33/335301}
	{\bibfield  {journal} {\bibinfo  {journal} {J. Phys. A: Math. Theor.}\
		}\textbf {\bibinfo {volume} {46}},\ \bibinfo {pages} {335301} (\bibinfo
		{year} {2013})}\BibitemShut {NoStop}%
	\bibitem [{\citenamefont {Peng}\ \emph {et~al.}(2014)\citenamefont {Peng},
		\citenamefont {Ren}, \citenamefont {Braak}, \citenamefont {Guo},
		\citenamefont {Ju}, \citenamefont {Zhang},\ and\ \citenamefont
		{Guo}}]{Peng_2014}%
	\BibitemOpen
	\bibfield  {author} {\bibinfo {author} {\bibfnamefont {J.}~\bibnamefont
			{Peng}}, \bibinfo {author} {\bibfnamefont {Z.}~\bibnamefont {Ren}}, \bibinfo
		{author} {\bibfnamefont {D.}~\bibnamefont {Braak}}, \bibinfo {author}
		{\bibfnamefont {G.}~\bibnamefont {Guo}}, \bibinfo {author} {\bibfnamefont
			{G.}~\bibnamefont {Ju}}, \bibinfo {author} {\bibfnamefont {X.}~\bibnamefont
			{Zhang}}, \ and\ \bibinfo {author} {\bibfnamefont {X.}~\bibnamefont {Guo}},\
	}\href {\doibase 10.1088/1751-8113/47/26/265303} {\bibfield  {journal}
		{\bibinfo  {journal} {J. Phys. A: Math. Theor.}\ }\textbf {\bibinfo {volume}
			{47}},\ \bibinfo {pages} {265303} (\bibinfo {year} {2014})}\BibitemShut
	{NoStop}%
	\bibitem [{\citenamefont {Sun}\ \emph {et~al.}(2020)\citenamefont {Sun},
		\citenamefont {Cong}, \citenamefont {Eckle}, \citenamefont {Ying},\ and\
		\citenamefont {Luo}}]{Sun2019}%
	\BibitemOpen
	\bibfield  {author} {\bibinfo {author} {\bibfnamefont {X.-M.}\ \bibnamefont
			{Sun}}, \bibinfo {author} {\bibfnamefont {L.}~\bibnamefont {Cong}}, \bibinfo
		{author} {\bibfnamefont {H.-P.}\ \bibnamefont {Eckle}}, \bibinfo {author}
		{\bibfnamefont {Z.-J.}\ \bibnamefont {Ying}}, \ and\ \bibinfo {author}
		{\bibfnamefont {H.-G.}\ \bibnamefont {Luo}},\ }\href {\doibase
		10.1103/PhysRevA.101.063832} {\bibfield  {journal} {\bibinfo  {journal}
			{Phys. Rev. A}\ }\textbf {\bibinfo {volume} {101}},\ \bibinfo {pages}
		{063832} (\bibinfo {year} {2020})}\BibitemShut {NoStop}%
	\bibitem [{\citenamefont {Irish}\ \emph {et~al.}(2005)\citenamefont {Irish},
		\citenamefont {Gea-Banacloche}, \citenamefont {Martin},\ and\ \citenamefont
		{Schwab}}]{Irish_2005}%
	\BibitemOpen
	\bibfield  {author} {\bibinfo {author} {\bibfnamefont {E.~K.}\ \bibnamefont
			{Irish}}, \bibinfo {author} {\bibfnamefont {J.}~\bibnamefont
			{Gea-Banacloche}}, \bibinfo {author} {\bibfnamefont {I.}~\bibnamefont
			{Martin}}, \ and\ \bibinfo {author} {\bibfnamefont {K.~C.}\ \bibnamefont
			{Schwab}},\ }\href {\doibase 10.1103/PhysRevB.72.195410} {\bibfield
		{journal} {\bibinfo  {journal} {Phys. Rev. B}\ }\textbf {\bibinfo {volume}
			{72}},\ \bibinfo {pages} {195410} (\bibinfo {year} {2005})}\BibitemShut
	{NoStop}%
	\bibitem [{\citenamefont {Irish}\ and\ \citenamefont
		{Gea-Banacloche}(2014)}]{Irish_2014}%
	\BibitemOpen
	\bibfield  {author} {\bibinfo {author} {\bibfnamefont {E.~K.}\ \bibnamefont
			{Irish}}\ and\ \bibinfo {author} {\bibfnamefont {J.}~\bibnamefont
			{Gea-Banacloche}},\ }\href {\doibase 10.1103/PhysRevB.89.085421} {\bibfield
		{journal} {\bibinfo  {journal} {Phys. Rev. B}\ }\textbf {\bibinfo {volume}
			{89}},\ \bibinfo {pages} {085421} (\bibinfo {year} {2014})}\BibitemShut
	{NoStop}%
	\bibitem [{\citenamefont {Sakurai}(2017)}]{J.J.Sakurai2017}%
	\BibitemOpen
	\bibfield  {author} {\bibinfo {author} {\bibfnamefont {J.~J.}\ \bibnamefont
			{Sakurai}},\ }\href
	{https://www.ebook.de/de/product/29832326/j_j_sakurai_jim_napolitano_modern_quantum_mechanics.html}
	{\emph {\bibinfo {title} {Modern Quantum Mechanics}}}\ (\bibinfo  {publisher}
	{Cambridge University Pr.},\ \bibinfo {year} {2017})\BibitemShut {NoStop}%
	\bibitem [{\citenamefont {Griffiths}(2018)}]{DavidJ.Griffiths2018}%
	\BibitemOpen
	\bibfield  {author} {\bibinfo {author} {\bibfnamefont {D.~J.}\ \bibnamefont
			{Griffiths}},\ }\href
	{https://www.ebook.de/de/product/31794885/david_j_griffiths_darrell_f_schroeter_introduction_to_quantum_mechanics.html}
	{\emph {\bibinfo {title} {Introduction to Quantum Mechanics}}}\ (\bibinfo
	{publisher} {Cambridge University Pr.},\ \bibinfo {year} {2018})\BibitemShut
	{NoStop}%
\end{thebibliography}


\end{document}